\documentclass[11pt,a4paper]{article}
\pdfoutput=1
\usepackage{jheppub}
\usepackage{amsmath}
\usepackage{wasysym}
\usepackage{verbatim}
\newcommand{\opluslhrim}{\mathbin{\rlap{$\Leftcircle$}{+}}}

\usepackage{appendix}
\usepackage{graphicx}
\usepackage{mathrsfs}
\usepackage{caption}
\usepackage{float}
\usepackage{subfig}

\newcommand{\e}{\epsilon}
\newcommand{\sss}{\scriptscriptstyle}

\newcommand{\s}{\star}

\newcommand{\p}{\partial}
\newcommand{\refb}[1]{(\ref{#1})}

\newcommand{\non}{\nonumber}

\newcommand{\be}[1]{ \begin{equation}\label{#1} }
\newcommand{\ee}{\end{equation}}
\newcommand{\bea}[1]{\begin{eqnarray}\label{#1} }
\newcommand{\eea}{\end{eqnarray}}
\newcommand{\bes}{\begin{subequations}}
\newcommand{\ees}{\end{subequations}}

\renewcommand{\a}{\alpha}

\newcommand{\D}{\Delta}

\renewcommand{\(}{\left(}
\renewcommand{\)}{\right)}

\renewcommand{\a}{\alpha}

\newcommand{\lb}{\Big[}
\newcommand{\rb}{\Big]}

\title{Field Theories on Null Manifolds}
\author[a]{Arjun Bagchi,}
\author[b,c]{Rudranil Basu,} \author[d,e]{Aditya Mehra,} 
\author[a,f]{and Poulami Nandi} \author{\\}

\affiliation[a]{Indian Institute of Technology Kanpur, Kalyanpur, Kanpur 208016. INDIA \\}

\affiliation[b]{Center for the Fundamental Laws of Nature, Harvard University, Cambridge, MA 02138, USA\\}

\affiliation[c]{Department of Physics, BITS-Pilani, K K Birla Goa Campus, NH-17B, Zuarinagar, Goa-403726. INDIA\\}

\affiliation[d]{International Institute of Physics, Federal University of Rio Grande do Norte,
Campus Universitario, Lagoa Nova, Natal, RN 59078-970. BRAZIL\\} 

\affiliation[e]{Max-Planck-Institut f$\ddot u$r Gravitationsphysik (Albert-Einstein-Institut), 14476 Golm, GERMANY\\} 

\affiliation[f]{Institute for Theoretical Physics, TU Wien, Wiedner Hauptstr. 8, A-1040 Vienna, AUSTRIA\\} 

\emailAdd{abagchi@iitk.ac.in, rbasu@g.harvard.edu, aditya.mehra@aei.mpg.de, poulamin@iitk.ac.in}

\abstract{We argue that generic field theories defined on null manifolds should have an emergent BMS or conformal Carrollian structure. We then focus on a simple interacting conformal Carrollian theory, viz. Carrollian scalar  electrodynamics. We look at weak (on-shell) and strong invariance (off-shell) of its equations of motion under conformal Carrollian symmetries. Helmholtz conditions are necessary and sufficient conditions for a set of equations to arise from a Lagrangian. We investigate whether the equations of motion of Carrollian scalar electrodynamics satisfy these conditions. Then we proposed an action for the electric sector of the theory. This action is the first example for an interacting conformal Carrollian Field Theory. The proposed action respects the finite and infinite conformal Carrollian symmetries in $d=4$. We calculate conserved charges corresponding to these finite and infinite symmetries and then rewrite the conserved charges in terms of the canonical variables. We finally compute the Poisson brackets for these charges and confirm that infinite Carrollian conformal algebra is satisfied at the level of charges.}

\preprint{}

\begin{document}

\maketitle

\newpage
\section{Introduction}
Null manifolds play starring roles on the stage of gravitational physics. Two of the most important classes of null manifolds are the null boundaries of asymptotically flat spacetime (henceforth referred to as $\mathcal{I^\pm}$) and the event horizons of generic black holes. 

In asymptotically flat spacetimes, the observables are the S-matrix elements and the asymptotic states of the S-matrix are defined at the null boundary $\mathcal{I^\pm}$. Null infinity is thus central to the formulation of quantum field theories in asymptotically flat spacetimes. The symmetries at $\mathcal{I^\pm}$ should also be encoded in the S-matrix. These symmetries at the boundary of spacetime are formally understood in terms of the asymptotic symmetry group (ASG). The ASG for asymptotically flat spacetimes in $d=4$ at null infinity turns out to be the infinite dimensional Bondi-Metzner-Sachs (BMS) group \cite{Bondi:1962px,Sachs:1962zza} instead of the expected Poincare group. This long-ignored BMS symmetry is now being used extensively to investigate the infra-red structure of gauge theories and gravity. The interplay between asymptotic symmetries, soft theorems and memory effects is opening up new avenues of investigation in quantum field theories and gravity \cite{Strominger:2017zoo}. 

Our other null surface of interest is the event horizon of a generic (non-extremal) black hole. This is of course the surface of no return for observers falling into the black hole. Intriguingly, on the event horizon, BMS symmetries have been discovered recently \cite{Hawking:2016msc,Hawking:2016sgy} (see \cite{Donnay:2015abr, Afshar:2016wfy, Penna:2017bdn, Carlip:2017xne, Donnay:2019jiz} for related work on symmetries of black hole event horizons) and this has been linked to a path that may lead to a possible resolution of the infamous black hole information loss paradox. 

The notion of holography first arose in the context of black hole thermodynamics and the fact that the entropy of a black hole is proportional to the area of its event horizon, rather than its volume. This lay the ground for the revolutionary proposal which linked gravity in $d$ dimensions with a theory without gravity living in $d-1$ dimension \cite{tHooft:1993dmi,Susskind:1994vu}. Holography today is mainly understood through the AdS/CFT correspondence \cite{Maldacena:1997re} for asymptotically AdS spacetimes. But we should also remember that black hole event horizons are inexorably linked to the idea of holography. 

In holography, the symmetries of the dual field theory arise from the asymptotic symmetries of the bulk theory. This was first exemplified by the work of Brown and Henneaux \cite{Brown:1986nw} for AdS$_3$, which can be thought of as a precursor of the Maldacena correspondence. Following the same line of thought, in the construction of a holographic correspondence for asymptotically flat spacetimes, BMS symmetries should play a central role and should dictate the symmetries of any putative dual theory \cite{Barnich:2010eb,Bagchi:2010eg}. Also, going beyond holography in asymptotically flat spacetimes, to the first idea of holography, a field theory dual to a generic black hole thought to be living on its event horizon, would also have to inherit the symmetries of the horizon. So in this case again, field theories with BMS symmetries arise as the main characters on the stage. In this paper, we will be interested in the construction and properties of such field theories. 

Null surfaces are characterised by degenerate metrics. We are interested in defining field theories on these surfaces. One of the ways to achieve this is to look at a systematic singular limit on a usual relativistic field theory on a space-like surface and give it an infinite boost. In terms of spacetime, this can be achieved by sending the speed of light $c$ in the field theory to zero \cite{Leblond65}. The underlying group then contracts from the Poincare group to what is known as the Carrollian group. We also want the field theory to live on the null surface and not move off it. So we will need that it be a massless theory. So we are led to a conformal version of the Carroll group, which has recently been shown to be isomorphic to the BMS group \cite{Duval:2014uva}. The above can be considered as a heuristic argument as to why there is a generic BMS symmetry associated to any field theory constructed on a null surface. 

The field theories that we are interested in are thus conformal field theories, defined on manifolds where the Riemannian structure has been replaced by a Carrollian structure. Since the Carrollian algebra is defined by a $c\to0$ contraction of the Poincare algebra, it is natural to construct the Conformal Carrollian Algebra (CCA), by the same contraction now applied to the relativistic conformal algebra \cite{Bagchi:2012cy}.  In this paper, we review this construction. An interesting aspect of our previous algebraic approach has been the construction of an  infinite dimensional lift for the CCA in arbitrary spacetime dimensions, which potentially makes it stronger than its relativistic parent \cite{Bagchi:2016bcd,Bagchi:2019xfx}. We also comment on this. We then look more geometrically at the construction of symmetries by considering conformal isometries of Carrollian manifolds and defining the symmetries for flat conformal Carrollian structures. We find that these symmetries form a general class of algebras characterised by an integer $N$ of which the $N=2$ member is the algebra that follows from the $c\to0$ contraction of the relativistic conformal algebra and is isomorphic to the BMS algebra. 

Having defined the background manifold of the field theories, we then go on to detailing the Carrollian field theories themselves. As with the algebra, one of the systematic ways of developing these field theories is to look at the $c \to 0$ limit of relativistic conformal field theories. This has been the path we have explored before in \cite{Bagchi:2016bcd,Bagchi:2019xfx}. Theories so obtained generically exhibit infinite dimensional BMS symmetry at the level of equations of motion. This led us to conjecture that infinite dimensional symmetry enhancements were a feature of the Carrollian version of any relativistic CFT \cite{Bagchi:2019xfx}. 


One of the drawbacks of this procedure of taking singular limits on relativistic CFTs was that there was no systematic way to do the same on the action and hence no prescription of how to build an action formulation for the contracted theory.{\footnote{See however \cite{Bergshoeff:2017btm} for action formulations of Carrollian theories.}} Although we don't directly solve this problem in our present paper, we construct our first example of an action of an interacting conformal Carrollian theory. In \cite{Basu:2018dub}, an action for Carrollian electromagnetism was put forward. In this paper, we examine Carrollian electrodynamics coupled to a massless scalar. We revisit our earlier formulation of finding symmetries of the equations of motion first and then put forward the notion of strong (off-shell) invariance of the equations of motion under certain symmetries, following \cite{Beisert:2017pnr,Beisert:2018zxs}. We show that Carrollian scalar electrodynamics displays strong invariance under the infinite conformal Carrollian symmetry in $d=4$, which is indicative of an action formulation of the theory. We do a further analysis of Helmholz conditions following \cite{Banerjee:2019axy} which form necessary and sufficient conditions for a set of equations of motion to arise from an action. We then put forward an action principle for the theory and look at the symmetries of the proposed action. We calculate the conserved charges corresponding to the finite and infinite Carrollian symmetries by using the Noether procedure similar to \cite{Basu:2018dub}. We finally compute the Poisson brackets to investigate the realisation of the corresponding algebra at the level of the conserved charges.

\section{Carrollian symmetries and conformal extensions}

\subsection{Ultra-relativistic limit of conformal symmetry}
The ultra-relativistic limit of conformal symmetries arises by performing an In{\"o}n{\"u}-Wigner contraction on the relativistic generators of conformal group. We implement this by taking the following limit on a $d$ dimensional Minkowski spacetime coordinates: 
\be{}\label{stc}
x_i\to x_i,~ t \to \e t,~ \e \to 0.
\ee
Here,  $i= 1, \ldots, (d-1)$. This implies taking the speed of light $c \to 0$. We apply this scaling and limit on the generators of relativistic conformal symmetry and regularise them. For example, the ultra-relativistic boost generator can be written from the relativistic boost generator in the following way:
\bea{}
&&\non J_{0i}=t \p_i +x_i \p_t \to \e t \p_i+\frac{1}{\e}x_i \p_t \, \Rightarrow B_i=\lim_{\e \to 0}\e J_{0i}=x_i \p_t.
\eea
In this way, the spacetime contraction \eqref{stc} on the generators of relativistic conformal symmetries give the following finite generators,
\bea{genur}
&&\non B_i=x_i \p_t, ~ J_{ij}=(x_i \p_j-x_j \p_i), ~ H=\p_t, ~ P_i=\p_i,\\
 &&D= (t \p_t+x_i \p_i),~ K= x_i x_i \p_t,~ K_j=2x_j(t\p_t+x_i\p_i)-(x_i x_i)\p_j.
\eea
These generators give rise to following non-vanishing Lie brackets,
\bea{}\label{algebra}
&&\non [J_{ij}, B_k ]=\delta_{k[j}B_{i]}, ~ [J_{ij}, P_k ]=\delta_{k[j}P_{i]},~ [J_{ij}, K_k ]=\delta_{k[j}K_{i]}, ~ [B_i,P_j]=-\delta_{ij}H,\\
&&\non  [B_i,K_j]=\delta_{ij}K,~ [D,K]=K,~[K,P_i]=-2B_i,~[K_i,P_j]=-2D\delta_{ij}-2J_{ij},\\ &&[H,K_i]=2B_i,~[D,H]=-H, ~[D,P_i]=-P_i,~[D,K_i]=K_i.
\eea
We will call the algebra in \eqref{algebra} the finite Carrollian Conformal Algebra (fCCA), consisting of boosts $B_i$, rotation generators $J_{ij}$ in $(d-1)$ spatial dimensions, time translation $H$, spatial translation $P_i$, dilatation $D$ and temporal and spatial part of special conformal transformation (SCT) $K,K_i$ respectively. The sub-algebra $\{J_{ij}, B_i, P_i, H\}$ forms the Carrollian algebra which is the $c\to0$ limit of the Poincare algebra. It is of interest to point out that the generators $\{J_{ij},P_i,D,K_i\}$ form a $so(d+1)$, the conformal algebra of $d-1$ dimensional Eucledian space, which is another subalgebra of fCCA $ = iso(d+1)$. 

In \cite{Bagchi:2016bcd, Basu:2018dub}, an injective homomorphism $X: \mathrm{fCCA} \hookrightarrow \mathrm{CCA} $ was constructed. Here CCA is an infinite dimensional lie-algebra with an infinite dimensional abelian ideal. The key to this infinite dimensional extension of the finite algebra found through the limit is the time translation generators with arbitrary spatial dependence:
\be{}
M_f=f(x^1, x^2, \dots ,x^{d-1})\p_t =: f(x) \p_t.
\ee
Here, $f(x)$ is an arbitrary polynomial of degree $\a$, valued in space of tensors transforming under $so(d-1)$ generated by $J_{ij}$. When  $f(x_i)=1,x_i,x_k x_k$ we obtain $M_f = H,B_i,K \in \mathrm{fCCA}$ respectively. The finite generators $\{B_i,J_{ij},H,P_i,D,K,K_i\}$ along with $M_f$ for arbitrary $f$ constitute the infinite dimensional CCA. As a Lie algebra, CCA is the semidirect sum: $so(d+1) \mathbin{\opluslhrim} \mathcal{A}$, of the conformal algebra of $d-1$ dimensional Eucledian space with the infinite dimensional abelian ideal $\mathcal{A}$ generated by elements like $M_f$.

\medskip

\noindent The action of the $so(d+1)$ part on $\mathcal{A}$, induced from fCCA (for $d\geq4$) is \cite{Basu:2018dub,Bagchi:2019xfx}:
\bea{infalgebra}
&&\non [P_i, M_f] =M_{\p_i f},\quad  [D,M_f] =M_h,~\text{where}~h=x_i \p_i f-f,\\
&&\non [K_i,M_f]= M_{\tilde{h}},~\text{where}~\tilde{h}=2x_i h-x_k x_k\p_i f,\\
&&[J_{ij},M_f]= M_g,~\text{where}~{g}=x_{[i}\p_{j]}f.
\eea
It is easy to see that $\mathcal{A}_{\alpha}$ are invariant subspaces of the adjoint action of $D$ and $J_{ij} \in so(d-1)$, whereas each $M_f \in \mathcal{A}_{\alpha}$ is an eigenvector of $D$ with eigenvalue $\alpha-1$. Although, not essential for the purpose of our paper, it would be useful for analysing ultra-relativistic conformal field theories in $d \geq 4$ to again organize each $\mathcal{A}_{\alpha}$ into irreducible representations of $so(d-1)$.

\subsection{The $N$-Conformal Carroll Algebra}
Having started with a purely algebraic approach to the construction of Conformal Carrollian symmetries, we now turn to a more geometric method. As is well know, relativistic conformal symmetry algebra is the algebra of conformal isometries in Minkowski space. Following the same route, we wish to build the conformal algebra on Carrollian manifolds. 

\medskip

\noindent 
Carrollian manifolds are non-Riemannian manifolds equipped with a degenerate rank-2 symmetric covariant tensor $g$ and a nowhere vanishing vector field $\xi$, defining the kernel of $g$: $i_{\xi} g = 0$. For our present purpose, we need not consider extra structures like connection on a Carroll manifold. With the present structure at our disposal, the most straightforward way to think about conformal transformations on Carroll manifolds, is to consider those transformations which preserve the doublet ($g, \xi$) upto conformal scaling factors:
\be{lie}
\pounds_X g = \mu g, \, \,  \pounds_X \xi = \lambda \xi. 
\ee
Here $\mu$ and $\lambda$ are a-priori unrelated real numbers. To draw a parallel with physics in Minkowski spacetime, we consider a $d$ dimensional `flat'-Carroll manifold with topology of $\mathbb{R} \times \mathbb{R}^{(d-1)}$, so that in the chart $(t, x^1,x^2, \dots , x^{d-1})$:
\be{}
g= \delta_{ij} dx^i \otimes dx^j , \, \xi = \p_t. 
\ee
For this simple case, one can readily find the solution to the differential equations \eqref{lie} as:
\begin{eqnarray} \label{Xsol}
X &=& f(x) \p_t + \omega^j{}_i x^i \p_j + p^i \p_i + \Delta (x^i \p_i - \frac{2\lambda}{\mu} t \p_t ) \non\\
&+& k_i \left( 2 x^i \left(-\frac{2\lambda}{\mu} t \p_t + x^k \p_k \right) -x^k x_k \delta^{ij}\p_j \right)
\end{eqnarray}
where $\omega_{ij}\, \mathrm{(anti-symmetric) }, \Delta, k^i$ are constants of integration and $f$ is an arbitrary function of $x^1, x^2, \dots , x^{d-1}$. Due to arbitrariness of the functions $f$, the Lie algebra of vector fields $X$ above is infinite dimensional. For regularity of the finite transformations, we choose 
\be{} -\frac{\mu}{\lambda} = N, \quad N \in \mathbb{Z}_+
\ee 
We will call this above infinite dimensional algebra the  $N$-conformal Carroll algebra (CCA$_N$). Comparing the generators of ultra-relativistic algebra as presented in the previous subsection, with \eqref{Xsol} above, we see that the algebra discussed earlier indeed is isomorphic to the present one when we consider $N=2$. 

The generators of CCA${}_N$ differ from \eqref{genur} in case of level $N$ Dilatation $D^{(N)}$ and spatial part of SCT $K_i^{(N)}$.
\be{}
D^{(N)}= \(\frac{2}{N}\;t \p_t+x_i \p_i\),~ K_j^{(N)}=2x_j\(\frac{2}{N}\;t\p_t+x_i\p_i\)-(x_i x_i)\p_j.
\ee
It is natural to ask how the algebras of CCA${}_N$ differ from each other for different $N$, or in particular from the case of $N=2$. To this end, we first note that the subalgebra $\{ J_{ij}, P_i, D^{(N)}, K_i^{(N)}\}$ is actually same for all $N$ and is isomorphic to $so(d+1)$.  However the action of $so(d+1)$ on $\mathcal{A}$ is different from \eqref{infalgebra} and has explicit dependence on $N$, therefore the concise way to write CCA${}_N$ would be as $so(d+1) \mathbin{\opluslhrim}_N \left(\bigoplus_{\alpha \geq 0} \mathcal{A}_{\alpha} \right)$. Explicitly, for $M_f \in \mathbb{I}_{\alpha}$ the $N$ dependent action is:  
\bea{}
&&\non   [D^{(N)},M_f] =(\alpha - \frac{2}{N})M_{f},\\
&& [K_i^{(N)},M_f]= M_{\tilde{h}_N} \in \mathcal{A}_{\alpha+1},~\text{where}~\tilde{h}_N=\left(\left(\alpha - \frac{2}{N}\right)x_i - x^jx_j \p_i \right) f.
\eea
For the purposes of this paper, we will be mainly focused on $d=4$. In $d=4$, we can rewrite the generator of infinite dimensional Abelian ideal more explicitly (as in \cite{Bagchi:2016bcd}, \cite{Bagchi:2019xfx}): 
\be{altinf}
M_f=M^{m_1,m_2,m_3} =x^{m_1}y^{m_2}z^{m_3}\p_t.
\ee
The explicit form of the algebra becomes
\bea{}
&&\non [P_x, M^{m_1,m_2,m_3}]=m_1 M^{m_1-1,m_2,m_3},\\
&&[J_{xy},M^{m_1,m_2,m_3}]=m_2 M^{m_1+1,m_2-1,m_3}-m_1 M^{m_1-1,m_2+1,m_3},\non\\
&&\non [D^{(N)},M^{m_1,m_2,m_3}]=\(m_1+m_2+m_3-\frac{2}{N}\) M^{m_1,m_2,m_3},\\
&& [K_x^{(N)},M^{m_1,m_2,m_3}]=\(m_1+2m_2+2m_3-\frac{4}{N}\) M^{m_1+1,m_2,m_3}\non\\&&\hspace{3.2cm} -m_1(M^{m_1-1,m_2+2,m_3}+M^{m_1-1,m_2,m_3+2}).
\eea
Similar expressions hold for other components of $P_i, K_i^{(N)}$ and $J_{ij}$.
 
\subsection{Representation Theory of CCA}
Now, we will review the transformation of different fields under different CCA generators. Classical fields can be considered to be tensors on the Carrollian manifold and the Carrollian generators of diffeomorphism act by Lie derivative on them. However here we will look into fields having different spin and scaling dimensions which will not necessarily be tensors. In order to understand this point we will describe here the scale-spin representation of infinite CCA.

\paragraph{Scale- Spin representation} In the scale- spin representation,  the states are labeled by their scaling dimension $\D$ and spin $j$ in $d=4$. 
\be{}
D |\Phi \rangle= \D |\Phi\rangle, ~~ J^2 |\Phi\rangle=j (j+1)|\Phi\rangle.
\ee
For $d>4$ this representation theory can be generalised. The fields then transform in irreducible representations of the rotation generating subalgebra $so(d-1)$ and also have scaling dimension $\D$.
Similar to the state operator correspondence in 2$d$ CFT, we postulate
\be{stateop}
\lim_{(x_i,t) \to (0,0)} \Phi(x_i,t) |0 \rangle = |\Phi \rangle.
\ee
The generators are promoted to operators and the brackets become commutators:
\be{stateops} 
[D ,\Phi(0,0)]= \D \Phi(0,0), ~ [J^2, \Phi(0,0)]=j (j+1)\Phi(0,0).
\ee
Also, 
\be{}
[H, \Phi(t,x) ]=\p_t \Phi(t,x),~~ [P_i, \Phi(t,x) ]=\p_i \Phi(t,x). 
\ee
Following \cite{Bagchi:2016bcd}, the Conformal Carrollian primaries are defined as
\be{}
[K_i, \Phi(0,0) ]=0,~[K, \Phi(0,0) ]=0,~[M_f,\Phi(0,0) ]=0\: \text{for the degree of polynomial } \a > 1. 
\ee
As the primaries are not eigenstates of Carrollian boosts $B_i$, we employ Jacobi identity to write the action of boosts on the primaries following \cite{Bagchi:2016bcd}. Then the most general transformation under boost becomes \cite{Bagchi:2019xfx}
\be{}
[B_k,\Phi(0,0)]= q \Phi_k+q^\prime \Phi \delta_{ik}.
\ee
Here, $\{\Phi,\Phi_k\}$ are primaries of different spins ($0,1$) and $\{q,q^\prime\}$ are some constants  to be determined from  the input from  dynamics.

Thus we are ready to write the transformation of spin 0 and spin $1$ primaries under different Conformal Carrollian generators. To write down the expression for the transformations of a generic field $\Phi$, we use the relation
\bea{} \delta_{\varepsilon}\Phi(t,x) = [\varepsilon Q,\Phi(t,x)], \eea
where $Q$ denotes the generators of the infinitesimal symmetry transformations and $\varepsilon$ is the symmetry parameter.
\paragraph{For scalars ($\varphi$):}
\bea{}\label{scalar}
\non\hspace{-.25cm}\text{Rotation: }\delta_{\omega} \phi(t,x)&&=\omega^{ij} (x_i \p_j-x_j \p_i ) \phi(t,x),\\
\non\hspace{-.25cm}\text{Boost: }\delta_{\sss B} \phi(t,x)&&=b^j(x_j\p_t \phi(t,x)+r \phi_j(t,x)),\\
\non\hspace{-.25cm}\text{Translation: }\delta_{p} \phi(t,x)&&=p^j \p_j \phi(t,x),\\ 
\non\hspace{-.25cm}\text{Dilatation: }\delta_{\Delta} \phi(t,x)&&= (\D+t\p_t+x_i \p_i) \phi(t,x),\\ 
\non\hspace{-.25cm}\text{SCT: } \delta_{k} \phi(t,x)
&&= k^j \lb(2\Delta x_j+2x_jt\partial_t+2x_i x_j \partial_i- x_i x_i \partial_j )\,\phi(t,x)+2t r \phi_j(t,x)\rb\\ 
\hspace{-.25cm}\text{Supertranslation:  }\delta_{M_f} \phi(t,x)&&= f(x)\p_t \phi(t,x)+r \phi_j(t,x)\p_j f(x).
\eea
\paragraph{For gauge fields $(B,A_i)$:}
\bea{}\label{vector}
\non\hspace{-.25cm}\text{Rotation: }\delta_{\omega} B(t,x)&=&\omega^{ij} (x_i \p_j-x_j \p_i ) B(t,x),\\
\non\hspace{-.25cm}          \delta_{\omega} A_l(t,x)&=&\omega^{ij} \lb(x_i \p_j-x_j \p_i ) A_l(t,x)+q^\prime\delta_{l[i}A_{j]}\rb,\, \\
\non\hspace{-.25cm}\text{Boost: }\delta_{\sss B} B(t,x)&=&b^j[x_j\p_t B(t,x)+qA_j(t,x)],\\
\non\hspace{-.25cm}         \delta_{B}A_l(t,x)&=&b^j[x_j\p_t A_l(t,x)+q^\prime \delta_{lj} B(t,x)],\,\\
\non\hspace{-.25cm}\text{Translation: }\delta_{p} B(t,x)&=&p^j \p_j B(t,x),\\
\non\hspace{-.25cm} \delta_{p} A_l(t,x)&=&p^j \p_j A_l(t,x),\,\\   
\non\hspace{-.25cm} \text{Dilatation: } \delta_{\Delta} B(t,x)&=& (\D^\prime+t\p_t+x_i \p_i) B(t,x),\,\\
\non\hspace{-.25cm} \delta_{\Delta} A_l(t,x)&=& (\D^\prime+t\p_t+x_i \p_i) A_l(t,x),\,\\ 
\non\hspace{-.25cm} \text{SCT: }\delta_{k} B(t,x)&=& k^j \lb(2\Delta^\prime x_j+2x_jt\partial_t+2x_i x_j \partial_i- x_i x_i \partial_j )\,B(t,x)+2t q A_j(t,x)\rb,\,\,\\ 
\non\hspace{-.25cm}  \delta_{k} A_l(t,x)&=& k^j \lb(2\Delta^\prime x_j+2x_jt\partial_t+2x_i x_j \partial_i- x_i x_i \partial_j )\rb\,A_l(t,x)\\
&&\non +2k_l x_j A_j(t,x)-2k_i x_l A_i(t,x)+2tq^\prime k_l B(t,x),\\ 
\non \hspace{-.25cm}\text{ST: }  \delta_{M_f} B(t,x)&=& f(x)\p_t B(t,x)+q A_i(t,x)\p_i f(x),\,\,\\
 \hspace{-.25cm} \delta_{M_f} A_l(t,x)&=& f(x)\p_t A_l(t,x)+q^\prime B(t,x)\p_l f(x).
\eea
For fields of higher spin, the action of the generators can be similarly constructed. Since in this paper, we will only be interested in fields of spins 0 and 1, we don't explicitly write the general case down. Another point to note in the above is that we have chosen the dilatation eigenvalue of $A_i$ and $B$ to be the same. This is because we anticipate both the fields as coming from a relativistic vector which would have a single dilatation eigenvalue. This, in a purely non-relativistic set-up, is not a strict requirement, but as the theories that we consider arise in the Carrollian limit of Lorentzian theories, we will stick to this assumption.

\subsection*{Representation theory for generalised level N:}
Now, we will discuss the difference in the representation theory for a generalised level-$N$ of CCA${}_N$. The action of the level-N Dilatation \eqref{stateop} and the conditions for the Conformal Carrollian primaries get modified to
\be{}
[D^{(N)},\Phi(0,0)]=\D\Phi(0,0),~~[K_i^{(N)}, \Phi(0,0) ]=0.
\ee
Subsequently,
\bea{}
&&U^{-1}D^{(N)}U=D^{(N)}+\frac{2}{N}tH+x_i P_i,\non\\
&&\non U^{-1}K_j^{(N)}U=K_j^{(N)}+\frac{4}{N}tB_j+2D^{(N)}x_j+2x_kJ_{jk}+\frac{4}{N}tx_j H+2x_jx_iP_i-x_ix_iP_j.~~~~
\eea{}
Using the above we can find the action of the generators on fields of different spin. 
\paragraph{For scalars ($\varphi$):}
\bea{}\label{modscalar}
&&\hspace{-2cm}\delta_{\Delta}^{(N)} \phi(t,x)= (\D+\frac{2}{N}t\p_t+x_i \p_i) \phi(t,x),\non\\ 
\non&&\hspace{-2cm} \delta_{k}^{(N)} \phi(t,x)
= k_j^{(N)} \lb(2\Delta x_j+\frac{4}{N}x_jt\partial_t+2x_i x_j \partial_i- x_i x_i \partial_j )\,\phi(t,x)\non\\&&\hspace{6.2cm}+\frac{4}{N}t r \phi_j(t,x)\rb.
\eea
\paragraph{For gauge fields $(B,A_i)$:}
\bea{}\label{modvector}\non
\delta_{\Delta}^{(N)} B(t,x)&=& (\D^\prime+\frac{2}{N}t\p_t+x_i \p_i) B(t,x),\,\\
\non \delta_{\Delta}^{(N)} A_l(t,x)&=& (\D^\prime+\frac{2}{N}t\p_t+x_i \p_i) A_l(t,x),\,\\ 
\non \delta_{k}^{(N)} B(t,x)&=& k_j^{(N)} \lb(2\Delta^\prime x_j+\frac{4}{N}x_jt\partial_t+2x_i x_j \partial_i- x_i x_i \partial_j )\,B(t,x)+\frac{4}{N}t q A_j(t,x)\rb,\,\,\\ 
\non \delta_{k}^{(N)} A_l(t,x)&=& k_j^{(N)} \lb(2\Delta^\prime x_j+\frac{4}{N}x_jt\partial_t+2x_i x_j \partial_i- x_i x_i \partial_j )\rb\,A_l(t,x)\\
&& +2k_l x_j A_j(t,x)-2k_i x_l A_i(t,x)+\frac{4}{N}tq^\prime k_l B(t,x).
\eea

\bigskip

\section{Interacting Carrollian Field Theory}
As advertised in the introduction, we shall be concerned with field theories that live on null surfaces and these are Conformal Carrollian field theories. We wish to build prototypical models of such field theories. In previous works \cite{Bagchi:2016bcd,Basu:2018dub,Bagchi:2019xfx}, Carrollian electrodynamics and Carrollian Yang-Mills theories have been constructed at the level of equations of motion. (See also \cite{Duval:2014uoa} for earlier work on Carrollian electrodynamics.)  In \cite{Bagchi:2019xfx}, Carrollian scalar and fermionic field theories, as well as (massless) matter fields coupled to Carrollian gauge theories were addressed. We wish to now build towards an action formulation of an interacting Carrollian field theory. The specific example we have in mind is Carrollian scalar electrodynamics. We will be inspired by previous construction of an action for Carrollian electrodynamics \cite{Basu:2018dub}.  

\subsection{Carrollian electrodynamics: a quick recap}\label{sec2}
We begin with a brief recapitulation of the the simplest Carrollian gauge theory, Carrollian electrodynamics.
The Carrollian theory of electrodynamics can be classified into two sectors based on the scaling of the gauge field components $(B,A_i)$. Here, $B$ is the temporal part of the gauge field, and $A_i$ is the spatial part. These two sectors are known as the Electric and Magnetic sector. In the Magnetic sector, as the name suggests,  magnetic field dominates over electric field and the gauge fields scale in a particular way: $B \to  \e B, \; A_i \to  A_i, \: \e \to 0$, while in the other sector, electric field dominates over magnetic field and the corresponding scaling is $B \to  B, \; \e A_i \to  A_i, \: \e \to 0$. The equations of motion for Carrollian Electrodynamics are
\bes \label{edeom}
\bea{} 
\label{urseleom} && \mbox{Electric Sector:} \quad \p_{i}\p_{i}B-\p_{i}\p_{t}A_{i}=0,~~ \p_{t}\p_{i}B-\p_{t}\p_{t}A_{i}=0, \\
\label{ursmageom} &&  \mbox{Magnetic Sector:} \quad \p_{i}\p_{t}A_{i}=0,~~ \p_{t}\p_{t}A_{i}=0.
\eea
\ees
The above formalism can be extended for its counterpart containing massless matter fields, viz. Carrollian electrodynamics with scalars and fermions. In the field theories with massless matter, we need to scale the scalar and fermion fields accordingly.  In \cite{Bagchi:2019xfx}, it was shown that each of these two theories could have Electric and Magnetic sectors within them. Each of the Electric and Magnetic sectors is further divided into more subsectors. The origin of the diverse subsectors lies within different possibilities of the scaling of constituent matter fields. In this paper, we are primarily interested in the Electric sector of Carrollian scalar electrodynamics. We will also briefly discuss the  Magnetic sector.

\subsection{Carrollian scalar electrodynamics}
We aim to realise Carrollian scalar electrodynamics as a limit of its parent relativistic theory. We begin our analysis writing the Lagrangian density of relativistic electrodynamics coupled to massless scalar field.
\begin{equation}
\mathcal{L}=-\frac{1}{4}F^{\mu \nu} F_{\mu \nu}+(D_\mu \phi)^{\dagger} (D^\mu \phi).
\end{equation}
Here, the field strength $F_{\mu \nu}$ is defined as 
$
F_{\mu \nu}=\partial_\mu A_\nu-\partial_\nu A_\mu.
$
The gauge field $A^\mu$ and the massless scalar field $\phi$ are charged under $U(1)$ gauge group. The gauge transformations are given as,
\be{}
A_\mu (x) \to A_\mu(x)+\p_\mu \a (x), \:\: \phi(x)\to e^{-i\a(x)}\phi(x).
\ee
Here, $\a$ is the gauge transformation parameter. The gauge covariant derivative is
$D_{\mu}=\partial_\mu + ie A_\mu
$ and
$e$ is the coupling parameter.
The relativistic equations of motion of massless  scalar field $\phi$ and  gauge field $A_{\mu}$ are given as,
 \bes \label{SED}
\begin{eqnarray}
&&D_\mu D^\mu \phi=0,\\
&&\partial_\mu F^{\mu \nu}- ie[\phi^\dagger( D^\nu \phi)- \phi (D^\nu \phi)^\dagger]=0.
\end{eqnarray} 
\ees
Our objective is to look into the Carrollian version of the above theory.  

\subsubsection{Electric sector}
We reach the Electric sector of Carrollian scalar electrodynamics by taking the following scaling on complex scalar field $\phi$ (also its conjugate $\phi^\star$) and the coupling parameter along with the scaling on the gauge fields:
\begin{equation}\label{sca}
 \phi \to \epsilon \phi, \; \phi^\star \to \epsilon \phi^\star, \; e \to \frac{e}{\epsilon},\;B \to B, \; A_i \to \epsilon A_i.
\end{equation}
We should point out that here unlike our previous construction of Carrollian scalar electrodynamics in \cite{Bagchi:2019xfx}, the electric charge also gets scaled. This will be important when we are looking to construct an action formulation later. 

Plugging the usual spacetime scaling for the Carrollian theories $x_i \to x_i,\, t \to \e t, \e \to 0$ and the field scalings  \eqref{sca} into the relativistic equations \eqref{SED} we recast the equations of motion for the Electric sector of Carrollian scalar electrodynamics as
\bes  \label{eqm}
\begin{eqnarray}\label{eqm1}
\partial_t \partial_t A_i- \partial_t \partial_i B&=&0,\\\label{eqm2}
\p_i(\p_t A_i-\p_i B)-ie\(\phi D_t^\star \phi^\star - \phi^\star D_t \phi\)&=&0,\\\label{eqm3}
\partial_t \partial_t \phi - e^2 B^2 \phi +2ie B \partial_t \phi +ie (\partial_t B)\phi\equiv D_tD_t \phi&=&0.
\end{eqnarray}
\ees
The gauge covariant derivative is 
 $D_t=\p_t+ieB$ and its conjugate is $D_t^\star=\p_t-ieB$. 
 
\subsubsection{Symmetries of the electric sector}
Now we will check the symmetries of the above equations of motion. We want to see if the theory posses finite Conformal Carrollian symmetries. Then we will investigate if there can be an infinite enhancement of these symmetries. The check of these symmetries in the present section is at the level of equations of motion.

Following \cite{Bagchi:2016bcd}, the symmetries of  a generic equation $\mathcal{D} \circ \Phi(t,x)= \mathcal{J}$ under a symmetry group $\mathcal{G}$  is realised in the following way: 
\be{gensym}
\delta_{\varepsilon} \left(\mathcal{D} \circ  \Phi(t,x)\right) = \mathcal{D} \circ \delta_{\varepsilon} \Phi(t,x)= \mathcal{D} \circ [\varepsilon^\a \mathcal{Q}_\alpha,\Phi(t,x)]=0.
\ee
Here the differential operator $\mathcal{D}$ acts on a field $\Phi(t,x)$ of arbitrary spin and  $\mathcal{J}$ is a source term in the equation of motion. The associated algebra of the symmetry group $\mathcal{G}$ has generators $\mathcal{Q}_\alpha$, and  $\varepsilon^\a$ is an infinitesimal parameter associated with the particular transformation. The appearance of zero on the RHS of \eqref{gensym} confirms the symmetry of the equation of motion under the particular transformation.

We implement the above method for checking the symmetries of equations \eqref{eqm}.  
Under  finite Carrollian transformations,
\begin{equation} \label{fin_wk_check}
\begin{split}
\delta_{D}[\text{\eqref{eqm1}}]&=0,~~\delta_{K_j}[\text{\eqref{eqm1}}]=0,\\
\delta_{D}[\text{\eqref{eqm2}}]&=ie\{(\Delta^\prime-\Delta)-(\Delta-1)\}  \(\phi (\partial_t \phi^\star)- \phi^\star (\partial_t \phi)\)+4 (\Delta-1)B \phi^{\star}\phi,\\
\delta_{K_j}[\text{\eqref{eqm2}}]&=2 (\Delta^\prime-1) \lb \p_i  (\partial_i B- \partial_t A_i)+2\p_iB-\p_t A_i\rb-4 (\Delta-1)\lb ie(\phi\p_t \phi^\star-\phi^\star\p_t \phi)\\
&-2e^2B\phi^\star\phi\rb-2ier\lb(\phi \p_t \phi_j^\star-\phi^\star \p_t \phi_j)+t(\phi\p_t\phi_j^\star+\phi_j\p_t\phi^\star-\phi^\star\p_t\phi_j-\phi_j^\star\p_t\phi) \rb,\\
\delta_{D}[\text{\eqref{eqm3}}]&=(\Delta^\prime-1)\lb ie B \partial_t \phi+ ie \partial_t(B  \phi)-2e^2B^2\phi \rb, \\
\delta_{K_j}[\text{\eqref{eqm3}}]&=2 x_j  \lb \{ (\Delta^\prime-1)+(\Delta-1)    \}ie\p_t (B \phi)-2(\Delta^\prime-1)e^2 B^2 \phi+(\Delta^\prime-1) ieB \p_t \phi \rb.
\end{split}
\end{equation}
The dilatation eigenvalues of the fields $(\phi, B, A_i)$ are all $\D=\D'=\frac{d-2}{2}=1$ \footnote{The Carrollian scaling dimension is obtained from the relativistic scaling dimension by taking UR scaling on the relativistic dilatation generator \cite{Bagchi:2016bcd}.  $$D^{\text{rel}}=x_i \p_t + t \p_i+\D^{\text{rel}}\xrightarrow{\text{UR limit}}x_i \p_t + t \p_i+\D^{\text{rel}}.$$ Hence, the form of the dilatation generator does not change after taking the limit. It suggests, the scaling dimension also remains unchanged $\D^{\text{rel}}=\frac{d-2}{2}=\D(\text{Carrollian})$. }. Also, the boost labels  of the scalar and gauge fields are $r=0,q=0,q^\prime=1$. Hence, the RHS of the above equations vanish indicating that the equations are invariant under finite Conformal Carrollian symmetries in $d=4$.
Under infinite Carrollian transformations,
\begin{eqnarray} \label{inf_wk_check}
 &&\delta_{M}[\p_t(\p_t A_i-\p_i B)]=0,~~\delta_{M}[D_tD_t\phi]=0,~~\delta_{M}[\eqref{eqm2}]=0. 
\end{eqnarray}
Here also, the RHS vanishes as $r=0,q=0,q^\prime=1$.
Hence, the theory possess finite and infinite Conformal Carrollian Symmetry at $d=4$, at the level of equations and motion.

\subsubsection{Magnetic sector}
The magnetic sector of scalar electrodynamics involves the scaling of the scalar and gauge field in the following manner:
\begin{equation}\label{sca}
 \phi \to \epsilon \phi, \; \phi^\star \to \epsilon \phi^\star, \; e \to \frac{e}{\epsilon},\;B \to \e B, \; A_i \to  A_i, \e \to 0.
\end{equation}
The scaling of the fields along with the ultra-relativistic spacetime scaling reduce the relativistic equations \eqref{sca} to
\be{} \label{eqmag}
\p_t\p_tA_i=0,~~\p_i\p_tA_i=0,~~\p_t\p_t\phi+e^2A_i^2\phi=0.
\ee
These are the equations of motion for the magnetic sector of Carrollian Scalar electrodynamics. Using methods described for the electric sector, it can be shown that these magnetic sector equations are also invariant under the infinite set of Conformal Carrollian generators. 

\section{Towards an action formulation}
Our main focus has been the study of symmetry properties of Carrollian scalar electrodynamics. Recent works  \cite{Duval:2014uoa, Bagchi:2016bcd, Bagchi:2019xfx}, various aspects of Carrollian field theories, including symmetry properties, have been studied at the level of equations of motion. The issue of symmetries and ultimately the question of quantum upliftment of these symmetries via Ward identities is most efficiently handled if we have an action describing the local field theory and the symmetry generators in question act point-wise on the fields, following the standard Noether procedure. In order to proceed in this direction, in this section we pass our equations of motion of Carrollian scalar electrodynamics for symmetries and for other dynamical structures \eqref{eqm1} and \eqref{eqmag} through a couple of stringent checks as follows. 

\subsection{Strong invariance}
In this section, we adopt a recent formulation of strong invariance of equations of motion following \cite{Beisert:2018zxs}. For systems which at best are described by equations of motion of the form $E(\Phi, \p \Phi) = 0$, the first step towards building of an action formulation would be the check of the validity of the on-shell condition:
\begin{eqnarray} \label{weakness}
\delta_{\star} E(\Phi, \p \Phi) \approx 0
\end{eqnarray}
for a symmetry generator $\delta_{\star}$. But, this is a rather weak condition as has been exemplified recently in \cite{Beisert:2018zxs}. In fact, a subset of these weak symmetries of equations of motion might not qualify as symmetries of action, had there been one. One then resorts to a stronger condition. In principle, the data of equations of motion is sufficient to check this strong condition of invariance. In order to understand this, one starts by designating the equation of motion `conjugate' \footnote{In case the theory has an action $S[\Phi, \p \Phi ]$, the notion of conjugacy is understood via the variation on the space of dynamical variables $\delta S[\Phi, \p \Phi ] =  \breve{\Phi}\, \delta \Phi $ which of course results in the equation of motion $\breve{\Phi} =0$.} to a degree of freedom $\Phi_I$ as $ \breve{\Phi}^I =0$ by variational principle. In terms of $\breve{\Phi}$, the condition \cite{Beisert:2018zxs} for a transformation to be a strong invariance of the theory is:
\begin{eqnarray} \label{strength}
\label{se} \delta_{\star}\breve{\Phi}^K = -\frac{\delta (\delta_{\star}\Phi_{I})}{\delta \Phi_K}~\breve{\Phi}^I.
\end{eqnarray}
This evidently is a stronger statement than \eqref{weakness} $\delta_{\star}\breve{\Phi}^K \approx 0$.

Symmetries having the property of being strong invariance are of major interest in light of recent developments in classical integrability \cite{Beisert:2017pnr, Beisert:2018zxs} in a couple of classical supersymmetric field theories. An infinite dimensional Yangian algebra describing these otherwise `hidden' non-local (and hence non-Noetherian symmetries) symmetries is responsible for integrable structure in the systems in concern. In this program of establishing Yangnian invariance, only those locally acting generators which are strong invariance for equations of motion are regarded as the `level-zero' generators. 

We do not venture into unravelling an infinite class of hidden symmetries for Carrollian field theories of our interest. However, we would like to see whether Carrollian scalar electrodynamics theory checked above for weak Carrollain conformal invariance \eqref{fin_wk_check}, \eqref{inf_wk_check} pass through the dynamically non-trivial and more stringent check of strong invariance as well. We will start off with a well understood Lorentz invariant theory which possesses an action formulation as a warm-up exercise.

\subsubsection{A warm-up exercise}
Before going to the theory of our interest, let us demonstrate the idea invariance through a straightforward example of a free theory. For that, we will consider a relativistic real massless scalar field $\varphi(x)$ in $d$-dimensional spacetime. The Lagrangian density is given by
\be{} \mathcal{L} = -\frac{1}{2}\p_{\mu}\varphi(x)\p^{\mu}\varphi(x),\ee
and the equation of motion is 
\be{sceom} \breve{\varphi}:= \p^{\mu}\p_{\mu}\varphi(x)=0.\ee
We wish to examine the weak and strong invariance of \eqref{sceom} under Poincare transformations. The action of Poincare generators on $\varphi$ is given by
\bea{}\label{dsde} \delta_{P_{\sigma} }\varphi =\p_{\sigma}\varphi,~~ ~\delta_{L_{\rho\sigma}}\varphi=(x_{\rho}\p_{\sigma}-x_{\sigma}\p_{\rho})\varphi.\eea
 For this theory, $Z_I = Z_K =\varphi$. Therefore, the weak condition becomes
\bea{}\label{dsd}\delta_{\star}\breve{\varphi}\thickapprox 0 ~~\text{where}~~\star=P_{\sigma},L_{\rho\sigma}.\eea
Under the action of translation on \eqref{dsd} becomes,
\bea{}\delta_{P_{\sigma}}\breve{\varphi} = \p^{\mu}\p_{\mu}(\p_{\sigma}\varphi)=\p_{\sigma}(\p^{\mu}\p_{\mu}\varphi)=\p_{\sigma}\breve{\varphi} \thickapprox 0. \eea
Similarly, under the action of Lorentz transformations, the weak condition is given by
\bea{}\delta_{L_{\rho\sigma}}\breve{\varphi} = \p^{\mu}\p_{\mu}(x_{\rho}\p_{\sigma}\varphi-x_{\sigma}\p_{\rho}\varphi)=(x_{\rho}\p_{\sigma}-x_{\sigma}\p_{\rho})\p^{\mu}\p_{\mu}\varphi\thickapprox 0. \eea
We see that \eqref{sceom} is weakly invariant under Poincare transformations \eqref{dsde}. Next step will be to check whether \eqref{sceom} satisfy strong condition or not. Under translation $P_{\sigma}$, the LHS of \eqref{se} is given by
\bea{}\delta_{P_{\sigma}}\breve{\varphi}(y) = \p^{\mu}\p_{\mu}(\p_{\sigma}\varphi(y))=\p_{\sigma}(\p^{\mu}\p_{\mu}\varphi(y))=\p_{\sigma}\breve{\varphi}(y).  \eea
The RHS of \eqref{se} is given by
\bea{} -\int d^{d}x~\frac{\delta (\delta_{P_{\sigma}}\varphi(x))}{\delta \varphi(y)}~\breve{\varphi}(x)=\int d^{d}x~\frac{\delta\varphi(x)}{\delta\varphi(y)}~\p_{\sigma}(\p^{\mu}\p_{\mu}\varphi(x))=\p_{\sigma}(\p^{\mu}\p_{\mu}\varphi(y)). \eea
We see that LHS = RHS is satisfied. Here, in LHS, $\p_{\sigma}= \frac{\p}{\p y^{\sigma}}$ whereas in RHS, we start by taking $\p_{\sigma}= \frac{\p}{\p x^{\sigma}}$ in the integral and the final result becomes dependent on the coordinate $y_{\sigma}$. Under Lorentz transformation, \eqref{se} becomes
\bea{}&&\text{LHS:}~~ \delta_{L_{\rho\sigma}}\breve{\varphi}(y) = \p^{\mu}\p_{\mu}[(y_{\rho}\p_{\sigma}-y_{\sigma}\p_{\rho})\varphi(y)]=(y_{\rho}\p_{\sigma}-y_{\sigma}\p_{\rho})\p^{\mu}\p_{\mu}\varphi(y).\non\\&& \text{RHS:}~~-\int d^{d}x~\frac{\delta (\delta_{L_{\rho\sigma}}\varphi(x))}{\delta \varphi(y)}~\breve{\varphi}(x)=\int d^{d}x~\frac{\delta\varphi(x)}{\delta\varphi(y)}~[(x_{\rho}\p_{\sigma}-x_{\sigma}\p_{\rho})\p^{\mu}\p_{\mu}\varphi(x)]\non\\&&\hspace{5.9cm}=(y_{\rho}\p_{\sigma}-y_{\sigma}\p_{\rho})\p^{\mu}\p_{\mu}\varphi(y).
 \eea 
The equation \eqref{sceom} is strongly invariant under Poincare transformations. Let us also check the strong invariance under scale transformation $(D)$. The action of dilatation on scalar field $\varphi$ is given by 
\bea{} \delta_{D}\varphi = (x^{\tau}\p_{\tau}+\tilde{\Delta})\varphi, \eea 
where $\tilde{\Delta}= \frac{d-2}{2}$ is the scaling dimension of field $\varphi$. Under $D$, we have
\bea{}&&\text{LHS:}~~ \delta_{D}\breve{\varphi}(y) = \p^{\mu}\p_{\mu}[(y^{\tau}\p_{\tau}+\Delta)\varphi(y)]=(y^{\tau}\p_{\tau}+\tilde{\Delta} +2)\p^{\mu}\p_{\mu}\varphi(y).\non\\&& \text{RHS:}~~-\int d^{d}x~\frac{\delta (\delta_{D}\varphi(x))}{\delta \varphi(y)}~\breve{\varphi}(x)=\int d^{d}x~\frac{\delta\varphi(x)}{\delta\varphi(y)}~[(x^{\tau}\p_{\tau}+d-\tilde{\Delta})\p^{\mu}\p_{\mu}\varphi(x)]\non\\&&\hspace{5.55cm}=(y^{\tau}\p_{\tau}+\tilde{\Delta} +2)\p^{\mu}\p_{\mu}\varphi(y).
 \eea 
We see that \eqref{sceom} is strongly invariant under scale transformation.
  
\subsubsection{Carrollian Electrodynamics}
We have explained the importance of having strong invariance in a field theory and its connection to Yangian symmetry. We have also explicitly looked for strong invariance in a relativistic scalar field theory which serves as an example for what we will be doing next. We will now move to the electric sector of Carrollian Electrodynamics and see the strong invariance under Carrollian symmetry. 
\subsubsection*{Strong invariance of Electric sector}
The equations in Electric limit can be written down as
\bes{}\label{cem}
\bea{}\label{cem1}\breve{A_i}=\p_t\p_t A_i -\p_{t}\p_i B=0,\\\label{cem2}
\breve{B}=\p_i\p_i B -\p_i \p_t A_i=0. \eea\ees
To see the strong invariance of \eqref{cem}, we will use the representation of Carrollian algebra where we define the action of Carrollian generators on gauge fields $B,A_i$ given in \eqref{vector}. This representation is determine by the set of values $(\D,\Sigma,q,q')$. For this sector, the values are given as
\bea{}\D=1,q=0,q'=1.\eea The expression for strong invariance \eqref{se} for this case becomes
\bes{}
\bea{}\label{ss}&& \delta_{\star}\breve{B} (t,x) = -\int d^{d-1}y\,dt'~\Big[\frac{\delta (\delta_{\star}A_{i}(t',y))}{\delta B(t,x)}~\breve{A}_i(t',y)+\frac{\delta (\delta_{\star}B(t',y))}{\delta B(t,x)}~\breve{B}(t',y)\Big],\\&&\label{ss1} 
\delta_{\star}\breve{A}_i (t,x) = -\int d^{d-1}y\,dt'~\Big[\frac{\delta (\delta_{\star}A_{j}(t',y))}{\delta A_{i}(t,x)}~\breve{A}_j(t',y)+\frac{\delta (\delta_{\star}B(t',y))}{\delta A_{i}(t,x)}~\breve{B}(t',y)\Big].
 \eea\ees
Under Carrollian generators $(B_l,D,K,K_l)$, the left hand side of \eqref{ss1} for \eqref{cem1} becomes
\bea{}&& \delta_{B_l}\breve{A_i}=x_l\p_t \breve{A_i},~\delta_{K}\breve{A_i}=x^2\p_t \breve{A_i},\non\\&&
\delta_{D}\breve{A_i}=[t\p_t +x_l\p_l +(\D+2)]\breve{A_i},\non\\&&
\delta_{K_l}\breve{A_i}=(2x_l t\p_t +2x_l x_k \p_k -x^2 \p_l +2\D x_l +4x_l)\breve{A_i}\non\\&&\hspace{2.3cm}+(2-2\D)\delta_{li} \p_t B +2x_k \delta_{li}\breve{A_k}-2x_i \breve{A_l}.
\eea
Similarly, the right hand side of \eqref{ss1} for \eqref{cem1} becomes
\bea{}&& \hspace{-.4cm} \delta_{B_l}\breve{A_i}=x_l\p_t \breve{A_i},~\delta_{K}\breve{A_i}=x^2\p_t \breve{A_i},\non\\&&\hspace{-.4cm}
\delta_{D}\breve{A_i}=[t\p_t +x_l\p_l +(1-\D+\delta_{ll})]\breve{A_i},\\&&\hspace{-.4cm}
\delta_{K_l}\breve{A_i}=(2x_l t\p_t +2x_l x_k \p_k -x^2 \p_l -2\D x_l +2x_l+2x_l \delta_{kk})\breve{A_i} +2x_k \delta_{li}\breve{A_k}-2x_i \breve{A_l}. \non
\eea
We will now move to the strong invariance of \eqref{cem2}. Under ultra-relativistic generators, the left hand side of \eqref{ss} for \eqref{cem2} becomes
\bea{}&& \delta_{B_l}\breve{B}=x_l\p_t \breve{B}-\breve{A_l},~\delta_{K}\breve{B}=x^2\p_t \breve{B}-2x_i \breve{A_i},\non\\&&
\delta_{D}\breve{B}=[t\p_t +x_l\p_l +(\D+2)]\breve{B},\non\\&&
\delta_{K_l}\breve{B}=(2x_l t\p_t +2x_l x_k \p_k -x^2 \p_l +2\D x_l +4x_l)\breve{B}\non\\&&\hspace{1.5cm}-2t\breve{A_l}+(4\D+2-2\delta_{ii})\p_l B -(2\D+4-2\delta_{ii})\p_t A_l.
\eea
Similarly, the right hand side of \eqref{ss} for \eqref{cem2} becomes
\bea{}&& \delta_{B_l}\breve{B}=x_l\p_t \breve{B}-\breve{A_l},~\delta_{K}\breve{B}=x^2\p_t \breve{B}-2x_i \breve{A_i},\non\\&&
\delta_{D}\breve{B}=[t\p_t +x_l\p_l +(1-\D+\delta_{ll})]\breve{B},\non\\&&
\delta_{K_l}\breve{B}=(2x_l t\p_t +2x_l x_k \p_k -x^2 \p_l -2\D x_l +2x_l+2x_l\delta_{kk})\breve{B}-2t\breve{A_l}.
\eea
From above analysis, we conclude that the equations \eqref{cem} are strongly invariant under Carrollian generators in $d=4$ dimensions.

Apart from these finite generators, the infinite Carroll `super-translations' parameterized by arbitrary space dependent functions $f(x)$ also satisfy the strong invariance conditions. For example, we have:
\begin{eqnarray} \label{sutra_lhs}
\delta_{M_f}\breve{B} = \p^2 (f\, \dot{B}) - \p_i \p_t \left( f \dot{A}_i + B \p_i f \right)
\end{eqnarray}
On the other hand, explicit calculations involving the transformations rules \eqref{ss} give the following 
\begin{eqnarray} \label{sutra_rhs}
&&-\int d^{d-1}y\,dt'~\Big[\frac{\delta (\delta_{M_f}A_{i}(t',y))}{\delta B(t,x)}~\breve{A}_i(t',y)+\frac{\delta (\delta_{M_f}B(t',y))}{\delta B(t,x)}~\breve{B}(t',y)\Big] \non \\
&&= -\p_i f \p_t\left(\dot{A}_i - \p_i B \right) + f \p_i \p_t \left( \p_i B - \dot{A}_i \right) .
\end{eqnarray}
Comparing \eqref{sutra_lhs} and \eqref{sutra_rhs} off-shell and repeating the same exercise for $\delta_{M_f}\breve{A}_i$, we conclude presence of strong invariance in $M_f$ generators.
\subsubsection*{Strong invariance of Magnetic sector}
After checking the strong invariance of the Electric sector for Carrollian electrodynamics, we move our attention to the magnetic one. We will see in the analysis that the Magnetic sector does not have strong invariance under the Carrollian generators. This statement is important as it also hints towards the lack of action formalism for this particular sector. We will postpone this discussion for the later part of the paper.
The equations of motion for the magnetic sector is given by,
\bes{}\label{magem}
\bea{}\label{magem1}\breve{B}=\p_i \p_t A_i=0,\\ \label{magem2}
\breve{A_j}=\p_t\p_t A_j=0. \eea\ees
The values of the constants for this sector are given by \be{valmag}\D=1,q=1,q'=0.\ee
We consider the first equation \eqref{magem1}. We plug in the action of Carrollian generators on the fields $B,A_j$ into equations \eqref{ss} and \eqref{ss1}. We also use the values of the constants \eqref{valmag}. Here, we explicitly show the case for dilation, boost and infinite dimensional generators. It can also be seen for the other generators. For the first equation, the LHS of \eqref{ss} becomes
\bea{}&&\hspace{-.5cm} \delta_{B_l}\breve{B}=x_l\p_t \breve{B}+\breve{A_l}, ~\delta_{D}\breve{B}=[t\p_t +x_l\p_l +(\D+2)]\breve{B},~\delta_{M_f}\breve{B}=f \p_t\breve{B}+(\p_i f)\breve{A_i}.~~
\eea
While, the RHS of  \eqref{magem1} becomes
\bea{}&& \delta_{B_l}\breve{B}=x_l\p_t \breve{B}, ~\delta_{D}\breve{B}=[t\p_t +x_l\p_l +(4-\D)]\breve{B},~\delta_{M_f}\breve{B}=f \p_t\breve{B}.
\eea
Similarly, for the second equation \eqref{magem2}, the LHS of \eqref{ss1} takes the following form,
\bea{}&& \delta_{B_l}\breve{A_j}=x_l\p_t \breve{A_j}, ~\delta_{D}\breve{A_j}=[t\p_t +x_l\p_l +(\D+2)]\breve{A_j},~\delta_{M_f}\breve{A_j}=f\p_t \breve{A_j}.
\eea
The RHS of  \eqref{ss1} gives,
\bea{}&& \delta_{B_l}\breve{A_j}=x_l\p_t \breve{A_j}-\delta_{lj}\breve{B}, ~\delta_{D}\breve{A_j}=[t\p_t +x_l\p_l +(4-\D)]\breve{A_j},~\delta_{M_f}\breve{A_j}=f\p_t \breve{A_j}-(\p_i f)\breve{B}.\hspace{1cm}\non
\eea
The LHS and the RHS match for neither of the equations in \eqref{magem} Hence, from the above analysis it is clear that the Magnetic sector of Carrollian electrodynamics does not have strong invariance under Carrollian generators.
\vspace{5mm}

\subsubsection{Carrollian scalar electrodynamics}\label{Carrollian scalar electrodynamics}
We will look at the strong invariance under Carrollian generators for equations of motion of Electric limit. 
The equations in Electric limit can be written down as
\bes{}\label{cems}
\bea{}\label{cems1}&&\breve{A_i}=\p_t\p_t A_i -\p_{t}\p_i B=0,\\&&\label{cems2}
\breve{B}=\p_i\p_i B -\p_i \p_t A_i+ie[(\phi\p_t\phi^{*}-\phi^{*}\p_t\phi)-2ieB\phi\phi^{*}]=0,\\&&
\breve{\phi}^{*}=\p_t\p_t\phi +ie(\p_t B)\phi +2ie B (\p_t \phi)-e^2 B^2 \phi=0.\label{cems3}
 \eea\ees
 To get the strong invariance for \eqref{cems} we will use the action of Carrollian generators on gauge fields \eqref{vector} and $\phi$ \eqref{scalar}. The set of values of the constants appearing in the representation theory are taken as 
 \bea{}\D'=1,q=0,q'=1,\D=1,r=0.\eea
We took $\D'$ for the gauge fields and $\D$ for the scalar field.
 The expression \eqref{se} for this theory becomes
 \bea{}\label{ses} \delta_{\star}\breve{Z}_K (t,x) = -\int d^{3}y~\frac{\delta (\delta_{\star}Z^{I}(t,y))}{\delta Z^{K}(t,x)}~\breve{Z}_I(t,y) . \eea
Here, $Z(t,x)=A_i,B,\phi^*,\phi$.
Under the Carrollian conformal generators, the left hand side of \eqref{ses} for \eqref{cems1} becomes
\bea{}&& \delta_{B_l}\breve{A_i}=x_l\p_t \breve{A_i},~\delta_{K}\breve{A_i}=x^2\p_t \breve{A_i},\non\\&&
\delta_{D}\breve{A_i}=[t\p_t +x_l\p_l +(\D'+2)]\breve{A_i},\non\\&&
\delta_{K_l}\breve{A_i}=(2x_l t\p_t +2x_l x_k \p_k -x^2 \p_l +2\D' x_l +4x_l)\breve{A_i}\non\\&&\hspace{2.3cm}+(2-2\D')\delta_{li} \p_t B +2x_k \delta_{li}\breve{A_k}-2x_i \breve{A_l},\non\\&&
\delta_{M_f}\breve{A_i}=f\p_t\breve{A_i}.
\eea
Similarly, the RHS of \eqref{ses} for \eqref{cems1} becomes
\bea{}&&\hspace{-.6cm} \delta_{B_l}\breve{A_i}=x_l\p_t \breve{A_i},~\delta_{K}\breve{A_i}=x^2\p_t \breve{A_i},\non\\&&\hspace{-.6cm}
\delta_{D}\breve{A_i}=[t\p_t +x_l\p_l +(1-\D'+\delta_{ll})]\breve{A_i},\non\\&&\hspace{-.6cm}
\delta_{K_l}\breve{A_i}=(2x_l t\p_t +2x_l x_k \p_k -x^2 \p_l -2\D' x_l +2x_l+2x_l \delta_{kk})\breve{A_i} +2x_k \delta_{li}\breve{A_k}-2x_i \breve{A_l},\non\\
&&\hspace{-.6cm}\delta_{M_f}\breve{A_i}=f\p_t\breve{A_i}.
\eea
We will now move to the strong invariance of \eqref{cems2}. Under $(B_l,D,K,K_l)$, LHS of \eqref{ses} for \eqref{cems2} becomes
\bea{}&& \delta_{B_l}\breve{B}=x_l\p_t \breve{B}-\breve{A_l},~\delta_{K}\breve{B}=x^2\p_t \breve{B}-2x_i \breve{A_i},\non\\&&
\delta_{D}\breve{B}=[t\p_t +x_l\p_l +(\D'+2)]\breve{B}+(2\D-\D'-1)[ie(\phi\p_t\phi^{*}-\phi^{*}\p_t\phi)]\non\\&&\hspace{1.4cm}+(2\D-2)[2e^2B\phi\phi^*],\non\\&&
\delta_{K_l}\breve{B}=(2x_l t\p_t +2x_l x_k \p_k -x^2 \p_l +2\D' x_l +4x_l)\breve{B}\non\\&&\hspace{1.5cm}-2t\breve{A_l}+(4\D'+2-2\delta_{ii})\p_l B -(2\D'+4-2\delta_{ii})\p_t A_l \non\\&&\hspace{1.5cm}+(4\D x_l -2\D'x_l -2x_l)[ie(\phi\p_t\phi^{*}-\phi^{*}\p_t\phi)]+(4\D-4)x_l[2e^2 B \phi\phi^*],\non\\&&
\delta_{M_f}\breve{B}=f\p_t\breve{B}-\breve{A_i}\p_if.
\eea
Similarly, the right hand side of \eqref{ses} for \eqref{cems2} becomes
\bea{}&& \delta_{B_l}\breve{B}=x_l\p_t \breve{B}-\breve{A_l},~K.\breve{B}=x^2\p_t \breve{B}-2x_i \breve{A_i},\non\\&&
\delta_{D}\breve{B}=[t\p_t +x_l\p_l +(1-\D'+\delta_{ll})]\breve{B},\non\\&&
\delta_{K_l}\breve{B}=(2x_l t\p_t +2x_l x_k \p_k -x^2 \p_l -2\D' x_l +2x_l+2x_l \delta_{kk})\breve{B}-2t\breve{A_l}\non\\&&
\delta_{M_f}\breve{B}=f\p_t\breve{B}-\breve{A_i}\p_if.
\eea
Finally, we will show the strong invariance of \eqref{cems3} under Carrollian symmetry. The left hand side of \eqref{ses} for \eqref{cems3} becomes
\bea{}&& \delta_{B_l}\breve{\phi}^*=x_l\p_t \breve{\phi}^*,~K.\breve{\phi}^*=x^2\p_t \breve{\phi}^*,\non\\&&
\delta_{D}\breve{\phi}^*=[t\p_t +x_l\p_l +(\D+2)]\breve{\phi}^*+(\D'-1)[ie(\p_t B)\phi +2ieB(\p_t \phi)-2e^2B^2\phi],\non\\&&
\delta_{K_l}\breve{\phi}^*=(2x_l t\p_t +2x_l x_k \p_k -x^2 \p_l +2\D x_l +4x_l)\breve{\phi}^*\non\\&&\hspace{1.5cm}+(2\D'-2)x_l[ie(\p_t B)\phi +2ieB(\p_t \phi)-2e^2B^2\phi],\non\\&&
\delta_{M_f}\breve{\phi^\star}=f\p_t\breve{\phi^\star}.
\eea
On similar note, the right hand side of \eqref{ses} for \eqref{cems3} becomes
\bea{}&& \delta_{B_l}\breve{\phi}^*=x_l\p_t \breve{\phi}^*,~K.\breve{\phi}^*=x^2\p_t\breve{\phi}^*,\non\\&&
\delta_{D}\breve{\phi}^*=[t\p_t +x_l\p_l +(1-\D +\delta_{ll})]\breve{\phi}^*,\non\\&&
\delta_{K_l}\breve{\phi}^*=(2x_l t\p_t +2x_l x_k \p_k -x^2 \p_l -2\D x_l +2x_l+2x_l \delta_{kk})\breve{\phi}^*,\non\\&&
\delta_{M_f}\breve{\phi^\star}=f\p_t\breve{\phi^\star}.
\eea
The equations \eqref{cems} are strongly invariant under Carrollian generators in $d=4$ dimensions. 

\medskip

\noindent Of course, since the EOM of the magnetic sector of Carrollian electrodynamics are not strongly invariant under the conformal Carrollian generators, the EOM with the inclusion of scalar would also not be invariant. We hence don't venture into the magnetic sector. 

\subsection{Helmholtz conditions for Carrollian Scalar ED}\label{helmCsed}
We have shown above that the Electric sector of Carrollian Scalar Electrodynamics enjoys strong invariance under infinite Carrollian Conformal generators in $d=4$. This is a strong hint towards the existence of an action formulation. In this sub-section, we provide further evidence for the existence of such a formulation. We will check if indeed the equations of motion in question can come as Euler Lagrange equation of an action. 

\medskip

\noindent The necessary and sufficient conditions to determine the ``inverse" problem of calculus of variation is given by the Helmholtz conditions \cite{Henneaux:1984ke, Morandi:1990su, Nigam:2016aa}. In \cite{Banerjee:2019axy}, the Helmholtz conditions were used for determining uniquely an action for the equations of motion of the magnetic sector of Galilean electrodynamics and violated for the corresponding magnetic sector. Here, we aim to check the Helmholtz condition for the electric sector of Carrollian scalar electrodynamics. For this, we write down the equations of motion for the Electric sector of Carrollian scalar electrodynamics in a slighly modified notation following \cite{Banerjee:2019axy}.
\bes\label{helmeom}
\bea{}
&&T_0=\breve{B}=\p_i \p_i B-\p_i \p_t A_i+ie\left[(\phi \p_t \phi^\star-\phi^\star  \p_t\phi)-2ie B\phi^\star \phi\right],\\
&&T_i=\breve{A_i}=\p_t \p_t A_i-\p_t \p_i B,\\
&&T_{\phi^\star}=\breve{\phi^\star}=-\p_t\p_t\phi-ie(\p_tB)\phi-2ieB(\p_t)\phi+e^2 B^2\phi.
\eea
\ees
The Helmholtz conditions for a set of equations $T_A$ involving fields and their derivatives $u^A, (u^A)_{ab}$ (assuming there are at most second order derivatives) are:  
\bes
\bea{}
&&\frac{\p T_A}{(\p u^B)_{ab}}=\frac{\p T_B}{(\p u^A)_{ab}}\\
&& \frac{\p T_A}{(\p u^B)_{a}}+\frac{\p T_B}{(\p u^A)_{a}}= 2 \p_b \frac{\p T_B}{(\p u^A)_{ab}}\\
&&\frac{\p T_A}{\p u^B}=\frac{\p T_B}{\p u^A}-\p_a \frac{\p T_B}{(\p u^A)_{a}}+\p_a \p_b \frac{\p T_B}{(\p u^A)_{ab}}
\eea
\ees
Here, the fields are described by $u^A$. Hence, the indices $A,B$ can take the values $0,i$ or $\phi^\star$ corresponding to each of the fields present in the theory $B,A_i,\phi$. The variation of the equations of motion $T_A$ are w.r.t the fields $u^A$ or their first and second derivatives $(u^A)_a$ and $(u^A)_{ab}$ respectively. The EOM of the electric sector of conformal Carrollian scalar electrodynamics that we have been interested in \eqref{helmeom} indeed do satisfy the above Helmholtz conditions. In the appendix \ref{ApA}, we give the details of this computation.

\medskip

\noindent Let us reemphasise an important point here. We chose to scale the electric charge in our formulation of Carrollian scalar ED (see \refb{sca}) as opposed to our earlier construction in \cite{Bagchi:2019xfx}. As mentioned earlier, this was to facilitate the formulation of an action principle. One can check that the equations of motion of our previous formulation of Carrollian scalar ED does not satisfy the Helmholz conditions and hence cannot arise from an action without the inclusion of additional fields. We discuss this in detail in appendix \ref{ApB}.

\section{Action for Carrollian scalar electrodynamics}
We discussed the concept of strong invariance in general. We have also seen that the field theories such as Carrollian electrodynamics and Carrollian scalar electrodynamics possess strong invariance. The strong invariance suggests that the Carrollian symmetries are present at the level of action for these theories. Our analysis of Helmholz conditions further reinforced this idea.  
In this section, we propose an action for the electric sector of Carrollian scalar electrodynamics. This will be the very first example of an action of an interacting field theory which respects Conformal Carrollian invariance in $d=4$. The equations of motion \eqref{eqm} will follow from the proposed action. 




\subsection{The proposed action}
Our proposal for the Lagrangian for the electric sector of conformal Carrollian scalar electrodynamics in $d$ dimensions is
\be{lag}
L=\int_{\Sigma}\, \(\frac{1}{2}\lb (\p_i B)^2+\dot{A}_i^2-2 \dot{B}\,\p \cdot A\rb -(D_t^\star \phi^\star) (D_t \phi)\)d^{d-1} x.
\ee
Here we have used the following convention \be{}
D_t \phi= \dot{\phi}+ie B \phi,~D_t^\star \phi^\star=\dot{\phi}^\star-ie B \phi^\star,~\dot{B}=\p_t B,~~\dot{A}_i=\p_t A_i,~\dot{\phi}=\p_t \phi,~ \p \cdot A=\p_i A_i.
\ee
The spatial slice $\Sigma$ has a topology ${\rm I\!R}^{d-1}$ with $t=$ constant. The metric on $\Sigma$ is $\delta_{ij}$.
The action is given by $S=\int dt L$. Varying the action we get 
\bes
\bea{}
&\p_t(\p_t A_i-\p_i B)=0,~D_t D_t \phi=0,~~D_t^\star D_t^\star \phi^\star=0,\\ 
& \p_i(\p_t A_i-\p_i B)-ie\(\phi D_t^\star \phi^\star - \phi^\star D_t \phi\)=0.
\eea
\ees
The equations of motion are the same as those obtained from the limiting approach \eqref{eqm}. We have assumed that at the boundary $\p \Sigma$, the variation of the fields are zero. Alternatively,
\be{}
\p_i B \delta B|_{\p \Sigma}=0,~\dot{B}\delta A_i|_{\p \Sigma}=0.
\ee
\subsection{Symmetries of the action}
Now, we will analyse the symmetries of the action. In a general relativistic set up, consider 
an infinitesimal transformation on a generic field $\phi(x^\mu)$, such that
\be{geneq}
\phi(x^\mu) \to \phi(x^\mu) + \alpha \delta \phi(x^\mu).
\ee
Here, $\alpha$ is the parameter of infinitesimal transformation. The Lagrangian $L$ is written in terms of the spatial integral of Lagrangian density $\mathcal L$. The corresponding action in $d$ dimensions $(\mu=0,1,\hdots d-1)$ is
 \be{}
 S=\int dt L= \int dt \:d^{d-1}x \mathcal{L} \:(\phi, \p_\mu \phi).
 \ee
This infinitesimal transformation \eqref{geneq} is called a symmetry if it leaves the equation of motion invariant. Alternatively, it means that the action remains invariant. Hence, the Lagrangian $\mathcal L$ is allowed to differ at-most by a total derivative term under  the transformation of the field in \eqref{geneq}.

As an example, consider  the relativistic scalar field Lagrangian density $\mathcal{L}=|\p_\mu \phi|^2-m^2 |\phi^2|$ and the infinitesimal spacetime translation
 $x^\mu \to x^\mu +\a^\mu $. The parameter of infinitesimal translation is $\alpha$. The relativistic scalar field $\phi$ transforms under translation as,  
$
\delta \phi(x)=\a^\mu \p_\mu \phi(x).
$
The Lagrangian density transforms as,
  \be{}
  \mathcal{L} \to  \mathcal{L}+ \delta \mathcal{L},~~\text{where,~} \delta \mathcal{L} = \a^\mu \p_\mu \mathcal{L}.
\end{equation}
The above equation shows that $\mathcal{L}$ transforms as a scalar under translation. It implies that the scalar field Lagrangian has translational symmetry.

Now, we will employ the similar technique to our case in the Carrollian set up. We will talk about the symmetries of the action, as described by the Lagrangian \eqref{lag}. For short hand notation we write,
\bea{}
\non L&=&\int_\Sigma \: d^{d-1}x \left( \frac{1}{2}\lb (\p_i B)^2+\dot{A}_i^2-2 \dot{B}\,\p \cdot A\rb -(D_t^\star \phi^\star) (D_t \phi)\right).
\eea
Varying the Lagrangian we get,
\begin{equation}{\label{varlag}}
\begin{split}
\delta {L}
&=\int_{\Sigma}\,{1 \over 2}\lb 2 (\p_i B) \p_i(\delta B)+ 2 (\p_t A_i) \p_t(\delta A_i) - 2(\p_i A_i) \p_t (\delta B)-2 (\p_t B) \p_i (\delta A_i)\rb\\
&~-\Big((D_t \delta \phi)^\s D_t \phi+D_t^\s \phi^\s D_t \delta \phi +ie(\phi D_t^\s \phi^\s-\phi^\s D_t\phi)\delta B\Big)\, d^{d-1}x.
\end{split}
\end{equation}
We will be using the variations of the fields $(\phi,B, A_i$)  with scaling dimensions $(\Delta,\Delta^\prime)$ under different Carrollian generators \eqref{scalar}, \eqref{vector}. We will plug them into \eqref{varlag} to check the symmetries of the Lagrangian. Thus, we obtain the variation of the Lagrangian
 under finite and infinite Carrollian generators,
 \bes{}
\begin{eqnarray}
\non\delta_D  L &=&\int_{\Sigma}\, d^{d-1}x \: \Big[ \p_t \left( \frac{1}{2}\lb (\p_i B)^2+\dot{A}_i^2-2 \dot{B}\,\p \cdot A\rb -(D_t^\star \phi^\star) (D_t \phi)\right)\\
\non &&~~-2(\Delta-1)(D_t \phi^\s D_t \phi)+(\D^\prime-1)\Big( 2e^2 B^2 \phi^\s \phi-ie B(\phi^\s \p_t \phi-\phi \p_t \phi^\s)\\
&& ~~+(\p_iB)^2+(\p_tA_i)^2-2(\p_i A_i)(\p_t B)\Big) \Big], ~~ \\
\non \delta_K  L &= &\int_{\Sigma}\, d^{d-1}x \:\Big[  \p_t\left( 2 k^j\, t \,\Big( \frac{1}{2}[ (\p_i B)^2+\dot{A}_i^2-2 \dot{B}\,\p \cdot A] -(D_t^\star \phi^\star) (D_t \phi)\Big)\right)\\
\non &&+ 2 \p_t (k^jA_j B) +k^j \{4(\D^\prime-1) x_j\mathcal{L} -4(\D^\prime-\D)((D_t \phi)^\star D_t \phi)\\
 &&-2ie x_j (\D^\prime-1)(\phi^\s D_t \phi-\phi (D_t \phi)^\s)-2(\D^\prime-1)A_j\p_t B \}\Big],\\
\delta_{M_f}  L &=& \int_{\Sigma}\, d^{d-1}x\: \partial_t  \left[(f(x) \left( \frac{1}{2}\lb (\p_i B)^2+\dot{A}_i^2-2 \dot{B}\,\p \cdot A\rb -(D_t^\star \phi^\star) (D_t \phi)\right)\right] .
\end{eqnarray}
\ees
The vanishing of the terms other than the total derivative terms constraint the scaling dimensions of scalar and gauge field to be $\D=1=\D^\prime$.  Hence, the finite and the infinite Conformal Carrollian symmetries are preserved at the level of action for scalar electrodynamics in $d=4$.

We have thus been able to provide an intrinsic description of Carrollian scalar electrodynamics without recourse to any relativistic theory and any limit. The equations of motion matches with our limiting construction in Sec~\ref{sec2}. As emphasised before, \eqref{lag} is the first example of a Lagrangian of an interacting theory with infinite Conformal Carrollian symmetry in $d=4$.

\section{Conserved charges and charge algebra}
In the previous section, we observed that the proposed Lagrangian \eqref{lag} for the interacting Carrollian theory of scalars and electrodynamics has infinite Conformal Carrollian symmetries in $d=4$. The Noether theorem states that for every continuous symmetry of the Lagrangian there exists a conserved current. The  charge associated with the current is constant in time. The aim of the current section is to find out the conserved charges associated with the infinite Conformal Carrollian symmetries. Later, we would like to understand if the Conformal Carrollian algebra is realised at the level of the charges.

\subsection{Conserved charges corresponding to the symmetries} 

We will use the Noether procedure to find the conserved charges corresponding to the symmetries  of the Lagrangian \eqref{lag}. First, let us chalk out a systematic procedure we would be employing heavily to find out the charges. A generic Lagrangian in $d$ spacetime dimensions can be written as a function of a field $\varphi$ and its derivatives $(\p_t \varphi, \p_i \varphi)$.
\be{}
L=L(\varphi,\p_t \varphi, \p_i \varphi ).
\ee
Consider the transformation of the field $\varphi \to \varphi +\delta_1 \varphi$. Varying the Lagrangian on-shell, we get the equation of motion for the generic field $\varphi$. Hence, schematically we can write,
\be{}
\delta L= \int \: \lb\p_t \Theta_t + \text{(EOM of } \varphi) \:\delta_1 \varphi \rb \:d^{d-1}x.
\ee
Here $\Theta$ is a function of $\varphi, \p_t \varphi, \p_i \varphi, \delta_1 \varphi$ containing all the total derivative terms of the variation of  Lagrangian. Let the field  satisfying the equation of motion be denoted by $\tilde{\varphi}$. Then the above equation gets modified to
\be{theta}
\delta L=\int d^{d-1}x\Big(\p_t \Theta_t  (\tilde{\varphi}, \p_t \tilde{\varphi}, \p_i \tilde{\varphi}, \delta_1 \tilde{\varphi})\Big) ~~: \text{on-shell.}
\ee
Next, consider an infinitesimal symmetry transformation of the field $\varphi \to \varphi +\delta_2 \varphi$. As it is a symmetry transformation, the Lagrangian is allowed to differ only by a total derivative term. Hence,
\be{alpha}
\delta L= \int d^{d-1}x\Big(\p_t \alpha_t  (\varphi, \p_t \varphi, \p_i \varphi, \delta_2 \varphi) \Big) ~~: \text{off-shell.}
\ee
The above equations \eqref{theta} and \eqref{alpha} must be equal onshell (for $\varphi=\tilde{\varphi}$) and also for a symmetry transformation (when $\delta_1=\delta_2$). Alternatively,
\bea{}\label{inter}
&\p_t \Theta_t =\p_t \alpha_t  ~: (\text{For } \varphi=\tilde{\varphi},\: \delta_1=\delta_2),~~
&\text{or, }\p_t(\Theta_t-\alpha_t)=0.
\eea
Hence, the conserved charge corresponding to the continuous symmetry transformation is given by
\be{eqQ}
Q=\int (\Theta_t-\alpha_t) \: \:d^{d-1}x.
\ee
We will be using the above formulation to find the conserved charges for the finite and infinite Conformal Carrollian generators. Below we show a few of the intermediate steps of the calculation.  
\subsubsection*{Result for $\Theta_t$ (on-shell):} Varying the Lagrangian \eqref{lag}, we first compute the temporal part of $\Theta$ according to \eqref{theta}
\be{}
 \Theta_t=\dot{A}_i \, \delta A_i-\p \cdot A\, \delta B- (D_t \phi)^\star\,\delta\phi-\delta\phi^\star \,D_t \phi.
\ee
\subsubsection*{Results for $\alpha_t$ (off-shell): }Next, we calculate the temporal part of $\alpha$ for each of the finite and infinite Conformal Carrollian generators following \eqref{alpha}. They are given by
\bea{}
\begin{split}
&\text{Rotation:}~~ \alpha_t(\omega)=0,\\
&\text{Translation:}~~ \alpha_t(p)=0,\\
&\text{Dilatation:}~~ \alpha_t(\Delta)=t\lb\frac{1}{2} ( \dot A_k )^2+\frac{1}{2} ( \p_k B )^2- \dot B \, \p \cdot A- (D_t \phi)^\star D_t \phi     \rb,\\
&\text{SCT:}~~ \alpha_t(k)=2x_i t \(\frac{1}{2} \dot{A}_l^2+ \frac{1}{2} (\p_l B)^2-\dot B\, \p \cdot A-  (D_t\phi)^\star (D_t \phi)   \)k_i+2k_i B A_i,\\
&\text{Supertranslation:}~~ \alpha_t(f)=\frac{1}{2}f \lb \dot{A}_i^2  +(\p_i B)^2-2 \dot{B} \, \p \cdot A-2 (D_t\phi)^\star (D_t \phi)-(\p_i\p_i f) B^2  \rb.
\end{split}
\eea
We have seen that for the Dilatation and Special conformal transformation (SCT), the Lagrangian is a total derivative, if the scaling dimensions of scalar and gauge fields are 1 ($\Delta=\Delta^\prime=1$).

Now we will be calculating the corresponding conserved charges from $\Theta_t$ and $\alpha_t$ that we have just computed. We will use the transformation of  scalar and gauge fields under different Carrollian generators as defined in \eqref{scalar} and \eqref{vector} and plug them back in  \eqref{inter} and \eqref{eqQ}. Thus, the corresponding conserved charges are written as
\subsubsection*{Conserved Charges:}
\bes{}
\label{char1}
\bea{}
&&\text{Rotation:}~Q(\omega)=\int d^{d-1}x ~\omega^{ij}   \lb \dot{A}_k\(x_{[i}\p_{j]}A_k+\delta_{k[i}A_{j]}\)-(\p \cdot A)\(x_{[i}\p_{j]}B\)\\
&&\hspace{6cm}\non~-(D_t \phi)^\star(x_{[i}\p_{j]}\phi)- (x_{[i}\p_{j]}\phi)^\star (D_t \phi)  \rb,\\
&&\text{Translation:}~Q(p)=\int d^{d-1}x ~ p^l\lb \dot{A}_i \p_l A_i- \p_l B \, \p \cdot A - (D_t \phi)^\star \p_l \phi-\p_l \phi^\star D_t\phi \rb,\hspace{1.5cm}\\
&&\text{Supertranslation:}~Q(f)=\frac{1}{2}\int d^{d-1}x ~ \lb f(x) (\dot{A}_i^2-\p_i B)^2-2f(x) (D_t \phi)^\star (D_t \phi) \rb,\\
&&\text{Dilatation:}\non~Q(\Delta)= \int d^3x ~ \lb \frac{1}{2} t \((\dot A_i)^2 -(\p_i B)^2  \)+ \dot{A}_i (x_j \p_j A_i)- (\p \cdot A)x_j \p_j B\\
\non &&\hspace{4.2cm} + (A_i \dot A_i- B\,\p \cdot A )-t (D_t \phi)^\star \dot \phi-t (\dot \phi)^\star D_t \phi + t (D_t \phi)^\star(D_t \phi)\\
&&\hspace{4.2cm}- (D_t \phi)^\star (x_j\p_j \phi )- (x_j\p_j \phi)^\star D_t \phi- (D_t \phi)^\star \phi-\phi^\star (D_t \phi )\rb,~~~~~ 
\\&& \non
\text{SCT:}~~Q(k)=\int d^3x~k^i~ \lb 2 x_j (\dot A_i A_j-\dot A_j A_i)+\dot A_j (2 x_i x_k \p_k-x_k x_k \p_i+2 x_i)A_j +t x_i (\dot A_j)^2\\&&\non\hspace{2.5cm}
+ ((2 \dot A_i B-x_i(\p_l B)^2)t-2A_i B)- (2x_i+2x_i x_k \p_k-x_k x_k\p_i)B (\p \cdot A)\\&&\non\hspace{2.5cm}
+ 2tx_i\((D_t \phi)^\star(D_t \phi)-(D_t \phi)^\star\dot \phi- (\dot \phi)^\star(D_t \phi)\)-2x_i \(  (D_t \phi)^\star \phi+ \phi^\star (D_t \phi) \)\\
&&\hspace{2.5cm}-(D_t \phi)^\star (2x_i x_k \p_k-x_k x_k \p_i)\phi-(2 x_i x_k \p_k-x_k x_k \p_i)\phi^\star (D_t \phi)\rb.
\eea\ees
The above charges are the conserved charges for the finite and infinite Conformal Carrollian generators for scalar electrodynamics.

\subsection{Realisation of charge algebra}
The previous subsection dealt with finding the conserved charges for the finite as well as infinite Conformal Carrollian symmetries possessed by the Lagrangian \eqref{lag}. Here we want to realise the CCA \eqref{algebra},\eqref{infalgebra} at the level of the conserved charges \eqref{char1}.

 The first part of the Noether theorem describes the conserved charges associated with the continuous symmetry transformation. The second part of the Noether theorem says that the algebra of the symmetry generators is realised at the level of the charges. Alternatively, the Poisson brackets of the charges in terms of canonical variables reproduce the  symmetry algebra. In this section, we will first rewrite the conserved charges in terms of the canonical variables. Next, we will  compute the Poisson brackets of them and then interpret the results.
 
Let us first identify the phase space variables in our interacting Conformal Carrollian scalar electrodynamics theory. The canonical momenta corresponding to the scalar $\phi$ and the gauge fields $B,A_i$ are given as,
\bes
\bea{canmom}
\pi_\phi&=&-(D_t \phi)^\star,~~\implies \dot \phi ^\star=-\pi_\phi+ie B \phi^\star,\\
\pi_\phi^\star&=&-(D_t \phi),~~\implies \dot \phi =-\pi_\phi^\star-ie B \phi,\\ \label{constrnt}
\pi_0&=&-\p \cdot A,\\
\pi_i&=&\dot A_i.
\eea
\ees
The third equation \eqref{constrnt} is a first class constraint in the theory. We will talk about its importance  soon.

Now, we will be plugging the canonical momenta $(\pi_\phi, \pi_\phi^\star, \pi_0, \pi_i)$ into the conserved charges \eqref{char1}. Thus we will be rewriting the conserved charges in terms of the phase space variables. The conserved charges are given as

\bes{}\label{charcan}
\bea{}
&&\non \text{Rotation:}~~Q(\omega)=\int d^{d-1}x ~\omega^{ij}   \lb \pi_k \(x_{[i}\p_{j]}A_k+\delta_{k[i}A_{j]}\)+\pi_0 \(x_{[i}\p_{j]}B\)\\
&&\hspace{7cm}+\pi_\phi (x_{[i}\p_{j]}\phi)+\pi_\phi^\star(x_{[i}\p_{j]}\phi)^\star \rb,\\&&
\text{Translation:}~~Q(p)=\int d^{d-1}x ~ p^l\lb \pi_i \p_l A_i+ \pi_0 \p_l B + \pi_\phi \p_l \phi+\ \pi_\phi^\star \p_l \phi^\star \rb,\\&&
\text{Supertranslation:}~~Q(f)=\frac{1}{2}\int d^{d-1}x ~ \lb f(x) (\pi_i-\p_i B)^2-2f(x) (\pi_\phi)^\star (\pi_\phi) \rb,\\&&
\text{Dilatation:}~~Q(\Delta)= \int d^3x ~ \lb \frac{1}{2} t \( ( \pi_i)^2 -(\p_i B)^2  \)+ \pi_i (x_j \p_j A_i)+ \pi_0 x_j \p_j B + A_i \pi_i \non\\&&\hspace{5cm}+B \pi_0-ie t B (\phi \pi_\phi-\phi^\star \pi_\phi^\star)+\pi_\phi x_j\p_j \phi+\pi_\phi^\star x_j\p_j \phi^\star\non\\&&\hspace{9cm}+\phi \pi_\phi+\phi^\star \pi_\phi^\star-t \pi_\phi \pi_\phi^\star\rb,\\&&
\text{SCT:}~~Q(k)=\int d^3x~k^i~ \lb 2 x_j (\pi_i A_j-\pi_j A_i)+\pi_j \{(2 x_i x_k \p_k-x_k x_k \p_i+2 x_i)A_j \non\\
\non &&\hspace{3cm}+t x_i \pi_j\}+\{t \(2 \pi_i B-x_i(\p_l B)^2\)-2A_i B\}+\pi_0 (2x_i+2x_i x_k \p_k\\
\non &&\hspace{3cm}-x_k x_k\p_i)B+ 2tx_i\(-\pi_\phi \pi_\phi^\star-ieB(\pi_\phi \phi-\pi_\phi^\star \phi^\star)\)+2x_i \(  \pi_\phi \phi+ \pi_\phi^\star \phi^\star \)\\
&&\hspace{3cm}+\pi_\phi (2x_i x_k \p_k-x_k x_k \p_i)\phi+(2 x_i x_k \p_k-x_k x_k \p_i)\phi^\star \pi_\phi^\star\rb.
\eea\ees
Now we are ready to analyse the Poisson brackets of these charges. As, there are only first class constraints (see \eqref{constrnt}) involved in the theory, we do not  need to calculate the Dirac brackets, but only the Poisson brackets \cite{Crnkovic:1987tz}.  At this point, let us pause for a moment to understand what it means in relation with CCA. Suppose, $a_1,a_2,a_3$ are generators of  CCA such that
\be{}
[a_1,a_2]=a_3.
\ee
As, $a_1,a_2,a_3$ generate symmetry transformations of the underlying theory, there are conserved charges corresponding to each of them. Let us describe them by $Q(a_1),Q(a_2),Q(a_3)$. Then, we  obtain the following computing the Poisson brackets of the charges 
\be{rlsn}
\{Q(a_1),Q(a_2)\}=Q(a_3).
\ee
Equation \eqref{rlsn} is explained as the realisation of the CCA algebra in terms of the charges.

We will now explore the Poisson brackets of the charges in \eqref{charcan}. First, we will calculate the Poisson brackets of the charges corresponding to the finite CCA generators. Consider the Poisson brackets between the dilatation charge $Q(\Delta)$ and translation charge $Q(p)$. It turns out
\be{al1}
\{Q(\Delta),Q(p)\}=-Q(p).
\ee
Here we have used the following relations,
\bea{}&&
\{\phi(x,t),\pi_\phi(y,t)\}=\delta^3(x-y),\non\\&&\{A_i(x,t),\pi_j(y,t)\}=\delta_{ij}\delta^3(x-y),\non\\&&\{B(x,t),\pi_0(y,t)\}=\delta^3(x-y).
\eea
The above Poisson brackets \eqref{al1} reflects the CCA bracket
\be{}
\lb D,P_i\rb=-P_i.
\ee
Similarly, consider another finite CCA bracket between rotation and translation
\be{al2}
\left[\omega^{ij}J_{ij},p^k P_k\right] =(\omega^{ij}p^k)(\delta_{kj}P_i-\delta_{ki}P_j)
=\tilde{p}^kP_k.
\ee
Here, $\omega^{ij},p_k$ are the parameters of rotation and translation. We rewrite 
$\tilde{p}^kP_k=(\omega^{ij}p_j)P_i-(\omega^{ij}p_i)P_j$. Now, in terms of the charges we obtain
\be{} \{Q(\omega),Q(p)\}=Q(\tilde p).
\ee
The above Poisson bracket is a direct manifestation of \eqref{al2}. Let us now, consider some other examples relating the infinite CCA generators.
\bea{}
&&\non [P_i, M_f] =M_{\p_i f},\quad  [D,M_f] =M_h,~\text{where}~h=x_i \p_i f-f,\\
&&\non [K_i,M_f]= M_{\tilde{h}},~\text{where}~\tilde{h}=2x_i h-x_k x_k\p_i f,\\
&&[J_{ij},M_f]= M_g,~\text{where}~{g}=x_{[i}\p_{j]}f.
\eea
In terms of the charges we obtain,
\bes{}
\bea{}
&&\{Q(p),Q(f)\}=Q(h^\prime),~~\text{where,~}h^\prime=p^l \p_l f,\\
&&\{Q(\Delta),Q(f)\}=Q( h),~~\text{where,~} h=x_j\p_j f-f,\\
&&\{Q(k),Q(f)\}=Q(\tilde h),~~\text{where,~}\tilde h=(2x_i x_k \p_k-x_kx_k\p_i-2x_i)f,\\
&&\{Q(\omega),Q(f)\}=Q(g),,~~\text{where,~}g=\omega^{ij}x_{[i}\p_{j]}f.
\eea
\ees
All of the above results confirm the CCA algebra being satisfied at the level of charges.

Now, we arrive at the most interesting CCA bracket which concerns both of the infinite generators. CCA suggests that the infinite supertranslations $M_f$ commute.
\be{al3}
[M_{f_1}, M_{f_2}] = 0.
\ee
In terms of the charges for the infinite supertranslations we obtain,
\be{}
\{Q(f_1),Q(f_2)\}=0.
\ee
There is no emergence of central terms in the Poisson brackets. The above relation is the direct realisation of \eqref{al3}.	

\section{Concluding remarks}

\subsection{A quick summary} 

Our principle achievement in this current work has been the construction, for the first time, of an action of an interacting Carrollian conformal field theory, viz. Carrollian electrodynamics coupled to a massless scalar.  

\medskip

\noindent We started from relativistic scalar electrodynamics and reached its Carrollian version through ultra-relativistic scaling. Carrollian scalar electrodynamics, like its closely related cousin Carrollian electrodynamics, shows two distinct sections depending on the scaling of the gauge field components. In the electric sector, the electric field dominates over the magnetic sector, and the opposite happens in the magnetic sector. We checked for symmetries of both the sectors of the theory at the level of equations of motion and found infinite dimensional conformal Carrollian invariance. 

\medskip

\noindent 
Next, we reviewed the notion of strong invariance following Beisert et al \cite{Beisert:2017pnr} and applied this to the case at hand. More precisely, we looked at weak (on-shell) and strong (off-shell) invariance of the equations of motion of Carrollian scalar electrodynamics under infinite Carrollian symmetries at the level of equations of motion. This was indicative of the existence of an action principle, which was further reemphasised by the fact that these equations satisfied the Helmholz conditions. 

\medskip

\noindent 
We then proposed our action for the electric sector of Carrollian scalar electrodynamics \eqref{lag}. Varying the action, we could reproduce the equations of motion obtained previously from the limiting procedure. Our action was invariant under infinite conformal Carrollian symmetry at $d=4$ and served as the first example of an action for an interacting Carrollian theory. 

\medskip

\noindent 
Next, we constructed the conserved charges corresponding to the finite and infinite symmetries employing the Noether procedure. As a final check, we showed that the infinite CCA is realised at the level of charges. We did not find any central extension term while calculating the Poisson brackets of these charges. 

\subsection{Discussions}

There are some obvious and some not so obvious generalisations that follow from the current work. Below is a list of them.

\begin{itemize}

\item{\em{Magnetic sector}:} It is clear from our discussions that the Magnetic sector of Carrollian electrodynamics and scalar electrodynamics cannot have an action formulation without the inclusion of extra fields. In \cite{Banerjee:2019axy}, the magnetic sector was analysed from the point of view of Helmholz conditions and it was found that the addition of an extra scalar or an extra vector does not help. We would like to find the minimal set of extra fields required to formulate an action principle for this theory. This is work in progress. 

\item{\em{Carrollian Yang-Mills}:} We would like an action formulation for Carrollian Yang-Mills theory, which was formulated earlier \cite{Bagchi:2016bcd,Bagchi:2019xfx} in terms of equations of motion. These theories are very interesting with many different sectors, all of which exhibit infinite dimensional BMS symmetries. Although important in its own right, this is going to be the stepping stone of our attempt to understand the limit of AdS/CFT as we go on to describe briefly below. 

\item{\em{Supersymmetric Carrollian theories}:} It is of great interest to formulate supersymmetric field theories on null backgrounds. These Super Carrollian field theories would possibly inherit an infinite dimensional symmetry as well, akin to the infinite dimensional lifts for super Galilean symmetries, found e.g. in \cite{Bagchi:2009ke}. We also briefly discussed Yangian structures in this paper. It is likely that these structures will naturally arise in Carrollian theories when we are looking at supersymmetric versions. Questions of integrability would be best addressed in this set-up. It would also be interesting to formulate these theories, and their background symmetries in more geometric terms, as we reviewed for the bosonic case early in our paper. 

\item{\em{Flat holography}:} One of our primary goals of this programme is to systematically build a concrete example for a field theory holographically dual  to asymptotically flat spacetimes. For this we wish to start with the best known example of the gauge gravity duality, viz. the original Maldacena correspondence between gravity on AdS$_5 \times S^5$ and $\mathcal{N}=4$ Super Yang-Mills (SYM) theory. We wish to take a systematic limit on both sides of the correspondence. Taking the limit on the bulk AdS radius going to infinity of course gets one to asymptotically flat spacetimes. This also corresponds to the ultra-relativistic contraction of the boundary field theory \cite{Bagchi:2012cy}. Taking this cue, it is natural to attempt the construction of the Carrollian version of  $\mathcal{N}=4$ SYM and the hope is that this will be the concrete example we are looking for. The prospect of infinite dimensional symmetries makes this venture even more tantalising.  

\item{\em{Tensionless strings}:} Our construction of Carrollian field theories have further implications for string theory in the tensionless limit, where BMS symmetries arise as residual symmetries on the worldsheet \cite{Bagchi:2013bga, Bagchi:2015nca} and relatedly the physics of the Hagedorn phase transitions \cite{Bagchi:2019cay}.  

\end{itemize}

\section*{Acknowledgements}
We would like to thank Kinjal Banerjee for discussions. This work is supported partially by the following grants: SERB/EMR/2016/008037 (AB), SERB/ERC/2017/000873 (AB), SERB/MTR/2017/000740 (AB), DST Inspire Faculty (RB), OPERA, BITS-Pilani (RB).
AB would like to thank BITS Goa for warm hospitality near the completion of this work. PN would like to thank Daniel Grumiller, Niels Obers, Gerben Oling for discussions.  PN would also like to thank Science and Engineering Research Board for the Overseas Visiting Doctoral Fellowship (SERB- OVDF) ODF/2018/000759 for financial support in Vienna during the course of this work. The warm hospitality of NORDITA at Stockholm, CEICO at Prague and Niels Bohr Institute at Copenhagen is also acknowledged. RB thanks hospitality of IIT Kanpur during various stages of the work. AM would like to thank Alexander von Humboldt Foundation for the Humboldt Research Fellowship for Postdoctoral Researchers for financial support in Germany during the course of this work. 
\newpage

\appendices
\section*{Appendices}
\section{Helmholtz Conditions for Carrollian Scalar ED}\label{ApA}
In this appendix, we elaborate on the Helmholtz conditions for Carrollian Scalar ED that we briefly explained in Sec \ref{helmCsed}
The Helmholtz conditions for a set of equations $T_A$ involving fields and their derivatives $u^A, (u^A)_{ab}$ are given by
\bes
\bea{}
&&\frac{\p T_A}{(\p u^B)_{ab}}=\frac{\p T_B}{(\p u^A)_{ab}}\\
&& \frac{\p T_A}{(\p u^B)_{a}}+\frac{\p T_B}{(\p u^A)_{a}}= 2 \p_b \frac{\p T_B}{(\p u^A)_{ab}}\\
&&\frac{\p T_A}{\p u^B}=\frac{\p T_B}{\p u^A}-\p_a \frac{\p T_B}{(\p u^A)_{a}}+\p_a \p_b \frac{\p T_B}{(\p u^A)_{ab}}
\eea
\ees
In our case, the equations of motion are cast as 
\bes
\bea{}
&&T_0=\p_i \p_i B-\p_i \p_t A_i+ie\left[(\phi \p_t \phi^\star-\phi^\star  \p_t\phi)-2ie B\phi^\star \phi\right],\\
&&T_i=\p_t \p_t A_i-\p_t \p_i B,\\
&&T_{\phi^\star}=-\p_t\p_t\phi-ie(\p_tB)\phi-2ieB(\p_t)\phi+e^2 B^2\phi.
\eea
\ees
The satisfaction of the conditions can be seen in the tables below. 

\flushleft {\em{First Helmholtz Condition}}
\begin{center}
\begin{tabular}{| c| c | c |}
\hline
LHS & RHS & $a,b$\\
\hline
\hspace{10mm}$\frac{\p T_0}{(\p A_i)_{ab}}=-1$\hspace{10mm} & \hspace{10mm}$\frac{\p T_i}{(\p B)_{ab}}=-1$ \hspace{10mm}& \hspace{10mm} $a=i,b=t$ \hspace{9mm} \\
\hspace{10mm}$\frac{\p T_{\phi^\star}}{(\p B)_{ab}}=0$\hspace{10mm} & \hspace{8mm}$\frac{\p T_0}{(\p \phi^\star)_{ab}}=0$ \hspace{10mm}& \hspace{10mm} for all $a,b$\hspace{9mm} \\
\hspace{10mm}$\frac{\p T_{\phi^\star}}{(\p A_i)_{ab}}=0$\hspace{10mm} & \hspace{10mm} $\frac{\p T_i}{(\p \phi^\star)_{ab}}=0$\hspace{10mm} &\hspace{10mm} for all $a,b$\hspace{9mm} \\
\hline
\end{tabular}
\end{center}

\bigskip

{\em{Second Helmholtz Condition}}
\begin{center}
\begin{tabular}{|c| c |c|}
\hline
LHS & RHS & $a,b$\\
\hline
$\frac{\p T_0}{(\p A_i)_{a}}+\frac{\p T_i}{(\p A_0)_{a}}$ & $2 \p_b \frac{\p T_i}{(\p B)_{ab}}$ & \\
$=0$ & $=0$ & for all $a,b$\\ 
\hline
\hspace{5mm}$\frac{\p T_i}{(\p B)_{a}}+\frac{\p T_0}{(\p A_i)_{a}}$\hspace{5mm} & \hspace{5mm}$2 \p_b \frac{\p T_0}{(\p A_i)_{ab}}$\hspace{5mm} & \hspace{5mm}\\
$=0$ & $=0$ & for all $a,b$\\ 
\hline
$\frac{\p T_{\phi^\star}}{(\p B)_{a}}+\frac{\p T_0}{(\p \phi^\star)_{a}}$ & $2 \p_b \frac{\p T_0}{(\p \phi^\star)_{ab}}$ & \\
$=-ie\phi+ie\phi =0$ & =0 & for ($a=t$, all $b$)\\ & & \hspace{5mm}hold also for other values of $a$\hspace{5mm} \\
 \hline
$\frac{\p T_{\phi^\star}}{(\p A_i)_{a}}+\frac{\p T_i}{(\p \phi^\star)_{a}}$ & $2 \p_b \frac{\p T_i}{(\p \phi^\star)_{ab}}$ & \\
$=0$ & $=0$ & for all $a, b$\\ 
\hline
\end{tabular}
\end{center}

\bigskip

{\em{Third Helmholtz condition}}
\begin{center}
\begin{tabular}{|c | c |c|}
\hline
LHS & RHS & $a,b$\\
\hline
\hspace{3mm}$\frac{\p T_\phi^\star}{\p B}$\hspace{3mm} & \hspace{3mm}$\frac{\p T_0}{\p \phi^\star}-\p_a \frac{\p T_0}{(\p \phi^\star)_{a}}+\p_a \p_b \frac{\p T_0}{(\p \phi^\star)_{ab}}$\hspace{3mm} & \\
\hspace{3mm}$=-2ie\p_t \phi+2 e^2 B\phi$\hspace{3mm} & \hspace{3mm}$=2 e^2 B\phi-2 ie\p_t\phi$ & for $a=t$, all $b$\hspace{3mm}\\
& & holds trivially for all other values \\
\hline
$\frac{\p T_0}{\p \phi^\star}$ & $\frac{\p \phi^\star}{\p B}-\p_a \frac{\p  \phi^\star}{(\p B)_{a}}+\p_a \p_b \frac{\p T_{ \phi^\star}}{(\p B)_{ab}}$ & \\
$=-2ie\p_t \phi+2 e^2 B\phi$ & $=2 e^2 B\phi-2 ie\p_t\phi$ & for $a=t$, all $b$\\
& & \hspace{5mm}holds trivially for all other values\\
\hline
$\frac{\p T_\phi^\star}{\p A_i}$ & $\frac{\p T_i}{\p \phi^\star}-\p_a \frac{\p T_i}{(\p \phi^\star)_{a}}+\p_a \p_b \frac{\p T_i}{(\p \phi^\star)_{ab}}$ & \\
$=0$ & $=0$ & for all $a,b$\\
\hline
$\frac{\p T_0}{\p A_i}$ & $\frac{\p T_i}{\p B}-\p_a \frac{\p T_i}{(\p B)_{a}}+\p_a \p_b \frac{\p T_i}{(\p A_i)_{ab}}$ & \\
$=0$ & $=0$ & for all $a, b$\\
\hline
$\frac{\p T_i}{\p B}$ & $\frac{\p T_0}{\p  A_i}-\p_a \frac{\p T_0}{(\p A_i)_{a}}+\p_a \p_b \frac{\p T_0}{(\p A_i)_{ab}}$ & \\
$=0$ & $=0$ & for all $a, b$\\
\hline
\end{tabular}
\end{center}

\newpage
\section{No action for Carrollian SED without scaling $e$}\label{ApB}
In \cite{Bagchi:2019xfx}, we analysed the symmetries of the electric sector of Carrollian scalar eletcrodynamics. The major difference  was that we did not scale the coupling parameter $e$ in the earlier approach. Also, we considered the complex scalar field $\varphi$ in terms of two real scalar fields $\{\varphi_1, \varphi_2\}$ such that $\varphi=\varphi_1+i \varphi_2$. The scaling involved was
\bea{}
B\to B,\;\; A_i \to \e A_i, \varphi_1 \to \e^{p_1} \varphi_1, \varphi_2 \to \e^{p_2} \varphi_2.
\eea
Here, $\{p_1,p_2\}$ are two arbitrary constants. We identified the allowed values of $\{p_1,p_2\}$ in our analysis by imposing consistency conditions. There were several allowed sectors with different values of $\{p_1,p_2\}$. The symmetries of the equations of motion for all of these sectors exhibited invariance under the infinite dimensional CCA. For our analysis here, let us choose an arbitrary sub sector $\{p_1=0,p_2=1\}$. The corresponding equations of motion are
\bes
\label{eqprev}
\bea{}
&&T_0=\p_i \p_i B-\p_i\p_tA_i+2e\left[(\varphi_2 \p_t \varphi_1-\varphi_1 \p_t \varphi_2)-e B^2 \varphi_1^2\right]=0,\\
&& T_i=\p_t\p_tA_j-\p_t\p_jB=0,\\
&& T_{\varphi_1}=\p_t\p_t\varphi_1=0,\\
&& T_{\varphi_2}= \p_t\p_t\varphi_2+e\left[ 2B \p_t \varphi_1+\varphi_1 \p_t B\right]=0.
\eea
\ees
The invariance of the equations of motion under infinite Carrollian Conformal generators in $d=4$ is essentially the condition of weak invariance. We would like to check here whether these equations of motion satisfy the Helmholtz conditions. For convenience, we are only showing the non vanishing values for each of the conditions.

\flushleft {\em{First Helmholtz Condition}}
\begin{center}
\begin{tabular}{| c| c | c |}
\hline
LHS & RHS & $a,b$\\
\hline
\hspace{10mm}$\frac{\p T_0}{(\p A_i)_{ab}}=-1$\hspace{10mm} & \hspace{10mm}$\frac{\p T_i}{(\p B)_{ab}}=-1$ \hspace{10mm}& \hspace{10mm} $a=i,b=t$ \hspace{9mm} \\

\hline
\end{tabular}
\end{center}
Rest of the values vanish trivially. \\
\vspace{2mm}

\bigskip

{\em{Second Helmholtz Condition}}
\begin{center}
\begin{tabular}{|c| c |c|}
\hline
LHS & RHS & $a,b$\\
\hline
\hspace{10mm}$\frac{\p T_0}{(\p \varphi_1)_{a}}+\frac{\p T_{\varphi_1}}{(\p B)_{a}}$\hspace{10mm} & \hspace{10mm}$2 \p_b \frac{\p T_i}{(\p B)_{ab}}$\hspace{10mm} & \\
$=2e\varphi_2$ & $=0$ & for $a=t$, all $b$\\ 
\hline
\hspace{10mm}$\frac{\p T_0}{(\p \varphi_2)_{a}}+\frac{\p T_{\varphi_2}}{(\p B)_{a}}$\hspace{10mm} & \hspace{10mm}$2 \p_b \frac{\p  T_{\varphi_2}}{(\p B)_{ab}}$\hspace{10mm} & \\
$=-2e\varphi_1+2e\varphi_1=0$ & $=0$ & for $a=t$, all $b$\\ 
\hline
\hspace{10mm}$\frac{\p T_{\varphi_2}}{(\p \varphi_1)_{a}}+\frac{\p T_{\varphi_1}}{(\p \varphi_2)_{a}}$\hspace{10mm} & \hspace{10mm}$2 \p_b \frac{\p T_{\varphi_1}}{(\p \varphi_2)_{ab}}$\hspace{10mm} & \\
$=2eB$ & $=0$ & for $a=t$, all $b$\\ 
\hline
\end{tabular}
\end{center}

\newpage

{\em{Third Helmholtz condition}}
\begin{center}
\begin{tabular}{|c | c |c|}
\hline
LHS & RHS & $a,b$\\
\hline
\hspace{3mm}$\frac{\p T_0}{\p \varphi_1}$\hspace{3mm} & \hspace{3mm}$\frac{\p  \varphi_1}{\p B}-\p_a \frac{\p  \varphi_1}{(\p B)_{a}}+\p_a \p_b \frac{\p T_ {\varphi_1}}{(\p B)_{ab}}$\hspace{3mm} & \\
\hspace{3mm}$=-4e^2 B^2 \varphi_1-2e\p_t \varphi_2$\hspace{3mm} & \hspace{3mm}$=0$ & for $a, b$\hspace{3mm}\\
\hline
\hspace{3mm}$\frac{\p T_0}{\p \varphi_2}$\hspace{3mm} & \hspace{3mm}$\frac{\p  \varphi_2}{\p B}-\p_a \frac{\p  \varphi_2}{(\p B)_{a}}+\p_a \p_b \frac{\p T_ {\varphi_2}}{(\p B)_{ab}}$\hspace{3mm} & \\
\hspace{3mm}$=2e\p_t \varphi_1$\hspace{3mm} & \hspace{3mm}$=-e\p_t\varphi_1$ & for $a=t, b$\hspace{3mm}\\
\hline
\end{tabular}
\end{center}

\bigskip

From the above tables, it is evident that the Helmholtz conditions are not satisfied and hence, we can not formulate an action which gives the equations of motion in \eqref{eqprev} without the addition of further fields into the system. This is the reason why we chose to scale the electric charge in our formulation of Carrollian scalar electrodynamics in this paper.

\newpage

\bibliographystyle{JHEP}
\bibliography{ccft}
\end{document}